\documentclass[11pt]{article}

\usepackage{amssymb}
\usepackage{bbold}
\usepackage{bm}
\usepackage{natbib}
\usepackage{color,url}
\usepackage{graphicx,verbatim,array,multicol,amsfonts,amsmath,subfigure}
\usepackage{fullpage}
\usepackage{color}
\usepackage{epsfig,latexsym,amssymb}
\usepackage{pdfpages}
\newtheorem{definition}{Definition}
\newtheorem{theorem}{Theorem}[section]
\newtheorem{remark}{Remark}[section]
\newtheorem{corollary}{Corollary}[section]

\newcommand\independent{\protect\mathpalette{\protect\independenT}{\perp}}
\def\independenT#1#2{\mathrel{\rlap{$#1#2$}\mkern2mu{#1#2}}}

\title{SMC-ABC methods for the estimation of stochastic simulation models of the limit order book}
\date{\today}
\author{Gareth W. Peters \\UCL, Department of Statistical Science, WC1E 6BT, London, UK\\ gareth.peters@ucl.ac.uk
\and Efstathios Panayi\\ UCL, Department of Computer Science, WC1E 6BT, London, UK\\ efstathios.panayi.10@ucl.uk
        \and  Francois Septier \\ Institute Mines-Telecom Lille, CRIStAL UMR CNRS 9189, France.}

\begin{document}
\maketitle

\section{Introduction to intractable likelihood models for high frequency financial market dynamics}

In this chapter we consider classes of models that have been recently developed for quantitative finance that involve modelling a highly complex multivariate, multi-attribute stochastic process known as the Limit Order Book (LOB). The LOB is the primary data structure recorded each day intra-daily for all assets on every electronic exchange in the world in which trading takes place. As such, it represents one of the most important fundamental structures to study from a stochastic process perspective if one wishes to characterize features of stochastic dynamics for price, volume, liquidity and other important attributes for a traded asset. In this paper we aim to adopt the model structure recently proposed by \cite{panayi2015stochastic}, which develops a stochastic model framework for the LOB of a given asset and to explain how to perform calibration of this stochastic model to real observed LOB data for a range of different assets. One can consider this class of problems as truly a setting in which both the likelihood is intractable to evaluate pointwise, but trivial to simulate, and in addition the amount of data is massive. This is a true example of big-data application as for each day and for each asset one can have anywhere between 100,000-500,000 data vectors for the calibration of the models.

The class of calibration techniques we will consider here involves a Bayesian ABC reformulation of the indirect inference framework developed under the multi-objective optimization formulation proposed recently by \cite{panayi2015stochastic}. To facilitate an equivalent comparison for the two frameworks, we also adopt a reformulation of the class of genetic stochastic search algorithms utilised by \cite{panayi2015stochastic}, known as NGSA-II \citep{deb2002fast}. We adapt this widely utilised stochastic genetic search algorithm from the multi-objective optimization algorithm literature to allow it to be utilised as a mutation kernal in a class of Sequential Monte Carlo Samplers (SMC Sampler) algorithms in the ABC context. We begin with the problem and model formulation, then we discuss the estimation frameworks and finish with some real data simulation results for equity data from a highly utilised pan-European secondary exchange formerly known as Chi-X, before it was recently aquired by BATS.

\subsection{Introduction to the LOB and related multi-queue simulation models}
The structure of a financial market dictates the form of interaction between buyers and sellers. Markets for financial securities generally operate either as quote driven markets, in which specialists (dealers) provide 2-way prices, or order driven markets, in which participants can trade directly with each other by expressing their trading interest through a central matching mechanism. The LOB is an example of the latter, and indicatively, the Helsinki, Hong Kong, Shenzhen, Swiss, Tokyo, Toronto, and Vancouver Stock Exchanges, together with Euronext and the Australian Securities Exchange operate as pure LOBs, while the New York Stock Exchange, NASDAQ, and the London Stock Exchange also operate a
hybrid LOB system \citep{gould2013limit}, with specialists operating for the less liquid securities. 

We will consider trading activity in the context of the LOB in this chapter. Market participants are typically allowed to place two types of orders on the venue: Limit orders, where they specify a price over which they are unwilling to buy (or a price under which they are unwilling to sell), and market orders, which are executed at the best available price. Market orders are executed immediately, provided there are orders of the same size on the opposite side of the book. Limit orders to buy (sell) are only executed if there is opposite trading interest in the order book at, or below (above), the specified limit price. If there is no such interest, the order is entered into the limit order book, where orders are displayed by price, then time priority. 

 \begin{figure}[ht]
 \begin{center}
 \includegraphics[width=0.75\textwidth]{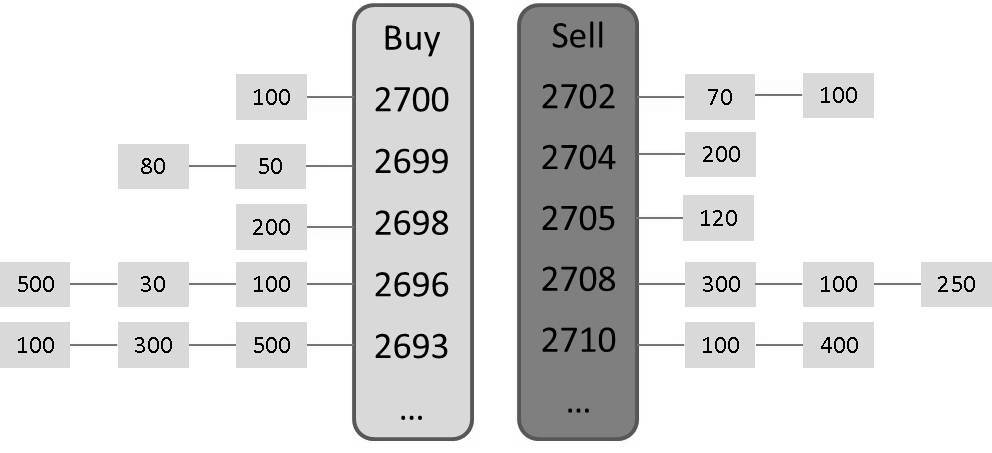}
 \caption{An example of the state of the Chi-X order book. The left hand side of the book is the buying interest, while the right hand side is the selling interest. The highest bid (order to buy) is for 100 shares at 2700 cents, while there are two lowest offers (orders to sell) for 70 and 100 shares at 2702. Orders are prioritised by price, then time.}
 \label{fig:OB}
 \end{center}
 \end{figure} 

Figure~\ref{fig:OB} shows an example snapshot of the order book for a particular stock, as traded on the Chi-X exchange, at a particular instance of time. A market order to buy 200 shares would result in 3 trades: 70 shares at 2702, another 100 shares at 2702 and the remaining 30 at 2704. A limit order to sell 300 shares at 2705, on the other hand, would not be executed immediately, as the highest order to buy is only at 2700 cents. It would instead enter the limit order book on the right hand side, second in priority at 2705 cents after the order for 120 shares which is already in the book.  

LOB simulation models aim to generate the trading interactions observed in such a LOB, and allow for a realistic description of the intra-day trading process. In particular, the models simulate the behaviour of individual market participants, usually based on the behaviour of various classes of real traders. The price of the traded financial asset is then determined from the limit and market orders submitted by these traders. Depending on the model, the instantaneous price is either considered to be the mid-point between the highest bid price and lowest ask price, or the last traded price. 

Because of the practical interest of modelling the intra-day dynamics of both stock prices and available volumes in the LOB, as well as the difficulty of traditional economic models based on rationality to reproduce these dynamics, there have been a multitude of research approaches attempting to address this gap. One one hand, there have been the zero-intelligence approaches (e.g. \cite{maslov2000simple}), which generally consist of a single, unsophisticated type of trading agent who submits orders randomly, possibly subject to (very few) constraints, like budget considerations. This minimum amount of discipline imposed by their actions is therefore often sufficient to reproduce some commonly observed features of financial markets, such as fat tails in the distribution of returns \citep{maslov2000simple}. Later models \citep{licalzi2003fundamentalists,fricke2013effects} also considered more realistic trading behaviour, where agents act based on their perceived value of the financial asset. However, \cite{licalzi2003fundamentalists} noted that these additional considerations regarding agent behaviour did not necessarily lead to a more realistic stock price output, and that it was likely the imposed market structure that had the largest effect on reproducing the price features observed. 

One other prominent strand of research pertains to the modelling of the LOB as a set of queues at each price and on both the buy (bid) and sell (ask) side. In the models of \cite{cont2010stochastic} and \cite{cont2013price}, a power law characterising the intensities of the limit order processes is found to be a good emprical fit. However, their assumption of independence between the activity at each level and for each event type is unlikely to hold in modern LOBs, due to the presence of algorithmic trading strategies, which induce different types of dependence structures. It is clear from observed empirical LOB data that non-trivial dependence structures are present, and as such, ignoring these features in the model formulation will result in inadequate representations of the stochastic LOB process being modelled.

In the LOB simulation model introduced in \cite{panayi2015stochastic}, they attempted to provide both a richer description of the LOB market structure and its constituent agents, but also consider the dependence between the trading activity at different levels. The main components of the proposed model are detailed in the following sections, before it is extended into an ABC posterior formulation for estimation and inference purposes. 

\section{Bayesian models with intractable likelihoods for high frequency financial market dynamics}
In this section we develop a new class of Bayesian models that can be utilised to study the dynamics of Limit Order Books (LOB) intra-daily. We start by presenting the stochastic multivariate order flow model, which we develop as a ``stochastic representative agent'' model framework. This is then reformulated as an intractable likelihood stochastic model and developed into an Approximate Bayesian Computational (posterior model). 

\subsection{Limit Order Book agent-based model}
\label{sec:lobmodel}
We consider the intra-day LOB activity in fixed intervals of time $\ldots,[ t-1,t ),[ t,t+1 ),\ldots$. For every interval $[ t,t+1 )$, we allow the total number of levels on the bid or ask sides of the LOB to be dynamically adjusted as the simulation evolves. These LOB levels are defined with respect to two reference prices, equal to $p_{t-1}^{b,1}$ and $p_{t-1}^{a,1}$, i.e. the price of the highest bid and lowest ask price at the start of the interval. We consider these reference prices to be constant throughout the interval $[ t-1,t)$ and thus, the levels on the bid side of the book are defined at integer number of ticks away from $p_{t-1}^{a,1}$, while the levels on the ask side of the book are defined at integer number of ticks away from $p_{t-1}^{b,1}$. 

This does not mean that we expect the best bid and ask prices to remain constant, just that we model the activity (i.e. limit order arrivals, cancellations and executions) according to the distance in ticks from these reference prices during this period. We note that it is of course possible that the volume at the best bid price is consumed during the interval, and that limit orders to sell are posted at this price, which would be considered at 0 ticks away from the reference price. To allow for this possibility, we actively model the activity at $-l_d+1, \ldots,0,\ldots, l_p$ ticks away from each reference price. Here, the $p$ subscript will refer to passive orders, i.e. orders which would not lead to immediate execution, if the reference prices remained constant and $d$ refers to direct, or aggressive orders, where it is again understood that they are aggressive with respect to the reference prices at the start of the period. Therefore, we actively model the activity at a total $l_t=l_p+l_d$ levels on the bid and ask.

We assume that activity that occurs further away is uncorrelated with the activity close to the top of the book, and therefore unlikely to have much of an impact on price evolution and the properties of the volume process. Therefore, the volume resting outside the actively modelled LOB levels ($-l_d+1, \ldots,0,\ldots, l_p$) on the bid and ask is assumed to remain unchanged until the agent interactions brings those levels inside the band of actively modelled levels, at which time they will again dynamically evolve. Such a set of assumptions is consistent with observed stylized features of all LOB for all modern electronic exchanges.

To present the details of the simulation framework, including the stochastic model components for each agent, i.e. liquidity providers and liquidity demanders, we first define the following notation:
\begin{enumerate}
\item $\bm{V}_t^{a}=(V_t^{a,-l_d+1},\ldots,V_t^{a,l_p})$ - the random vector for the number of orders resting at each level on the ask side at time $t$ at the actively modelled levels of the LOB at time $t$;
\item $\bm{N}_t^{LO,a}=(N_t^{LO,a,-l_d+1},\ldots,N_t^{LO,a,l_p})$ - the random vector for the number of limit orders entering the limit order book on the ask side at each level in the interval $[t-1,t)$;
\item $\bm{N}_t^{C,a}=(N_t^{C,a,1},\ldots,N_t^{C,a,l_t})$ - the random vector for the number of limit orders cancelled on the ask side in the interval $[t-1,t)$;
\item $N_t^{MO,a}$ - the random variable for the number of market orders submitted by liquidity demanders in the interval $[t-1,t)$.
\end{enumerate}

We consider the processes for limit orders and market orders, as well as cancellations to be linked to the behaviour of real market participants in the LOB. In the following, we model the aggregation of the activity of 2 classes of liquidity motivated agents, namely liquidity providers and liquidity demanders. As we model LOB activity in discrete time intervals, we process the aggregate activity at the end of each time interval in the following order:
\begin{enumerate}
\item Limit order arrivals - passive - by the liquidity provider agent;
\item Limit order arrivals - aggressive or direct - by the liquidity provider agent;
\item Cancellations by the liquidity provider agent;
\item Market orders by the liquidity demander agent.
\end{enumerate}

The rationale for this ordering is that the vast majority of limit order submissions and cancellations are typically accounted for by the activity of high-frequency traders, and many resting orders are cancelled before slower traders can execute against them. In addition, such an ordering allows us to condition on the state of the LOB, so that we do not have more cancellations at a particular level than the orders resting at that level. This is generally appropriate, as the time interval we consider can be made as small as desired for a given simulation. 

\subsubsection{Stochastic agent representation: liquidity providers and demanders}
We assume liquidity providers are responsible for all market-making behaviour (i.e. limit order submissions and cancellations on both the bid and ask side of the LOB). After liquidity is posted to the LOB, liquidity seeking market participants, such as mutual funds using some execution algorithm, can take advantage of the resting volume with market orders. For market makers, achieving a balance between volume executed on the bid and the ask side can be profitable. However, there is also the risk of adverse selection, i.e. trading against a trader with superior information. This may lead to losses if, e.g. a trader posts multiple market orders that consume the volume on several levels of the LOB. The risk of adverse selection as a result of asymmetric information is one of the basic tenets of market microstructure theory \cite{o1995market}. To reduce this risk, market makers cancel and resubmit orders at different prices and/or different sizes.

\begin{definition}[\textbf{Limit order submission process for the liquidity provider agent}]
Consider the limit order submission process of the liquidity provider agent to include both passive and aggressive limit orders on the bid and ask sides of the book, which are assumed to have the following stochastic model structure:
\begin{enumerate}
\item{Let the multivariate path-space random matrix $\bm{N}_{1:T}^{LO,k}\in \mathbb{N}_+^{l_t \times T}$ be constructed from random vectors for the numbers of limit order placements $\bm{N}_{1:T}^{LO,k} = \left(\bm{N}_1^{LO,k},\bm{N}_2^{LO,k}, \ldots,\bm{N}_T^{LO,k}\right)$. Furthermore, assume these random vectors for the number of orders at each level at time $t$ are each conditionally dependent on a latent stochastic process for the intensity at which the limit orders arrive, given by the random matrix $\bm{\Lambda}^{LO,k}_{1:T} \in \mathbb{R}_+^{l_t \times T}$ and on the path-space by $\bm{\Lambda}^{LO,k}_{1:T} = \left(\bm{\Lambda}_1^{LO,k},\bm{\Lambda}_2^{LO,k}, \ldots,\bm{\Lambda}_T^{LO,k}\right)$. In the following, $k \in \left\{a,b\right\}$ indicates the respective process on the ask and bid side.
}
\item{Assume the conditional independence property for the random vectors given by,
\begin{equation}
\left[\bm{N}_{s}^{LO,k}| \bm{\Lambda}^{LO,k}_{s}\right] \independent \left[\bm{N}_{t}^{LO,k}|\bm{\Lambda}^{LO,k}_{t} \right], \;\; \forall s \neq t,\;\; s,t \in \left\{1,2,\ldots,T\right\}.
\end{equation}
}
\item{For each time interval $[t-1,t)$ from the start of trading on the day, let the random vector for the number of new limit orders placed in each actively modelled level of the limit order book, i.e. the price points corresponding to ticks $(-l_d+1,\ldots,0,1,\ldots,l_p)$, be denoted by $\bm{N}_t^{LO,k}=(N_t^{LO,k,-l_d+1},\ldots,N_t^{LO,k,l_p})$, and assume that these random vectors satisfy the conditional independence property
\begin{equation}
\left[N_t^{LO,k,s}| \Lambda^{LO,k,s}_{t}\right] \independent \left[N_t^{LO,k,q}| \Lambda^{LO,k,q}_{t} \right], \;\; \forall s \neq q,\;\; s,q \in \left\{-l_d+1,\ldots,0,1,\ldots,l_p\right\}.
\end{equation}
}
\item{Assume the random vector $\bm{N}_t^{LO,k} \in \mathbb{N}_+^{l_t}$ is distributed according to a multivariate generalized Cox process with conditional distribution $\bm{N}_t^{LO,k} \sim \mathcal{GCP}\left(\bm{\lambda}^{LO,k}_t\right)$ given by 
\begin{equation} \label{EqnCntsDepLO}
\resizebox{.9\hsize}{!}{$
\mathbb{P}\text{r}\left(\left. N_t^{LO,k,-l_d+1} = n_1,\ldots,N_t^{LO,k,l_p} = n_{l_t}
\right|\bm{\Lambda}^{LO,k}_t = \bm{\lambda}^{LO,k}_t \right) = \prod_{s=-l_d+1}^{l_p} \frac{\left(\lambda_t^{LO,k,s}\right)^{n_s}}{n_s!}\,\exp\left[-\lambda_t^{LO,k,s}\right].
$}
\end{equation}
}
\item{Assume the independence property for random vectors of latent intensities unconditionally according to
\begin{equation}
\bm{\Lambda}_{s}^{LO,k} \independent \bm{\Lambda}_{t}^{LO,k}, \;\; \forall s \neq t,\;\; s,t \in \left\{1,2,\ldots,T\right\}.
\end{equation} 
}
\item{Assume that the intensity random vector $\bm{\Lambda}^{LO,k}_t \in \mathbb{R}_+^{l_t}$ is obtained through an element-wise transformation of the random vector $\bm{\Gamma}^{LO,k}_t \in \mathbb{R}^{l_t}$, where for each element we have the mapping 
\begin{equation} \label{EqnTransform}
\Lambda_t^{LO,k,s} = \mu_0^{LO,k,s} F\left(\Gamma_t^{LO,k,s}\right),
\end{equation}
where we have $s \in \left\{-l_d+1,\ldots,l_p\right\}$, baseline intensity parameters $\left\{\mu_0^{LO,k,s}\right\} \in \mathbb{R}_+$ and a strictly monotonic mapping $F:\mathbb{R} \mapsto [0,1]$.
}
\item{Assume the random vector $\bm{\Gamma}^{LO,k}_t \in \mathbb{R}$ is distributed according to a multivariate skew-t distribution $\bm{\Gamma}^{LO,k}_t \sim MSt(\bm{m}^k,\bm{\beta}^k,\nu^k,\Sigma^k)$ with location parameter vector $\bm{m}^k \in \mathbb{R}^{l_t}$, skewness parameter vector $\bm{\beta}^k \in \mathbb{R}^{l_t}$, degrees of freedom parameter $\nu^k \in \mathbb{N}_+$ and $l_t \times l_t$ covariance matrix $\Sigma^k$. Hence, $\bm{\Gamma}^{LO,k}_t$ has density function 

\begin{equation}
\resizebox{.9\hsize}{!}{$
f_{\bm{\Gamma}^{LO,k}_t}\left(\bm{\gamma}_t; \bm{m}^k,\bm{\beta}^k,\nu^k,\Sigma^k\right)= \frac{cK_{\frac{\nu^k+l_t}{2}}\left( \sqrt{(\nu^k+Q(\bm{\gamma}_t,\bm{m}^k))\left[\bm{\beta}^k\right]^T\left[\Sigma^k\right]^{-1}\bm{\beta}^k }\right)
\exp{(\bm{\gamma}_t-\bm{m}^k)}^T\left[\Sigma^k\right]^{-1}\bm{\beta}^k}
{\left( \sqrt{(\nu^k+Q(\bm{\gamma}_t,\bm{m}^k))\left[\bm{\beta}^k\right]^T\left[\Sigma^k\right]^{-1}\bm{\beta}^k} \right)^{-\frac{\nu^k+l_t}{2}} \left( 1+\frac{Q(\bm{\gamma}_t,\bm{m}^k)}{\nu^k} \right)^{\frac{\nu^k+l_t}{2}}}, 
$}
\end{equation}

where $K_v(z)$ is a modified Bessel function of the second kind given by
\begin{equation}\label{eq:bessel}
K_{v}(z)=\frac{1}{2}\int_{0}^{\infty} y^{v-1}e^{-\frac{z}{2}(y+y^{-1})}dy,
\end{equation}
and $c$ is a normalisation constant. We also define the function $Q(\cdot,\cdot)$ as follows:
\begin{align}
\label{eq:Q}
Q(\bm{\gamma}_t,\bm{m}^k)=(\bm{\gamma}_t-\bm{m}^k)^T \left[\Sigma^k\right]^{-1} (\bm{\gamma}_t-\bm{m}^k).
\end{align}
This model admits skew-t marginals and a skew-t copula, see \cite{smith2012modelling} for details. Importantly, this stochastic model admits the following scale mixture representation,
\begin{equation} \label{EqnSimGam}
\bm{\Gamma}^{LO,k}_t \stackrel{d}{=} \bm{m}^k + \bm{\beta}^k W + \sqrt{W}\bm{Z},
\end{equation}
with Inverse-Gamma random variable $W \sim IGa\left(\frac{\nu^k}{2},\frac{\nu^k}{2}\right)$ and independent Gaussian random vector $\bm{Z} \sim N\left(\bm{0},\Sigma^k\right)$.
}
\item{Assume that for every element $N_t^{LO,k,s}$ of order counts from the random vector $\bm{N}_t^{LO,k}$, there is a corresponding random vector $\bm{O}_t^{LO,k,s} \in \mathbb{N}_+^{N_t^{LO,k,s}}$ of order sizes. We assume that the element $O_{i,t}^{LO,k,s}, i \in \left \{ 1,\ldots, N_t^{LO,k,s}\right \} $ is distributed as $O_{i,t}^{LO,k,s} \sim H(\cdot)$. Furthermore, we assume that order sizes are unconditionally independent $O_{i,t}^{LO,k,s} \independent O_{i',t}^{LO,k,s}$ for $i \neq i'$, $s \neq s'$ and $t \neq t'$.
}
\end{enumerate}
\end{definition}

\begin{remark}
Under this proposed model for market maker liquidity activity, the number of limit orders placed by the liquidity providers in the market has an appropriate dynamic intensity structure that can evolve intra-daily to reflect the changing nature of liquidity provided by market makers throughout the trading day. In addition, the number of limit orders placed at each level of the bid and ask also allow for the model to capture the observed dependence structures in order placements in each level of the bid and ask regularly seen in empirical analysis of high-frequency LOB data. The dependence structure utilised is based on a skew-t copula which allows non-exchangeability of the stochastic intensity on the bid and ask at each level of the book, as well as asymmetry in the tail dependence features. This means that when large movements by market makers to replenish liquidity after a liquidity drought occurs intra-daily, such as after a large market order execution, the model can produce such replenishment on just the bid or just the ask depending on the situation. It does not automatically replenish both sides of the book equally likely, as would occur under a standard t-copula structure as opposed to a skew-t copula structure used in this model.
\end{remark}

We now define the second component of the liquidity provider agents, namely the cancellation process. The cancellation process has the same stochastic process model specification as the limit order submission process above, including a skew-t dependence structure between the stochastic intensities at each LOB level on the bid and ask. We therefore only specify the differences unique to the cancellation process relative to the order placement model definition in the below specification, to avoid repetition.

\begin{definition}[\textbf{Limit order cancellation process for liquidity provider agent}]
Consider the limit order cancellation process of the liquidity provider agent to have an identically specified stochastic model structure as the limit order submissions. The exception to this pertains to the assumption that the number of cancelled orders in each interval at each level is right-truncated at the total number of orders at that level.
\begin{enumerate}
\item{As for submissions, we assume for cancellations a multivariate path-space random matrix $\bm{N}_{1:T}^{C,k}\in \mathbb{N}_+^{l_t \times T}$ constructed from random vectors for the number of cancelled orders given by $\bm{N}_{1:T}^{C,k} = \left(\bm{N}_1^{C,k},\bm{N}_2^{C,k}, \ldots,\bm{N}_T^{C,k}\right)$. Furthermore, assume for these random vectors for the number of cancelled orders at each of the $l_t$ levels, the latent stochastic process for the intensity is given by the random matrix $\bm{\Lambda}^{C,k}_{1:T} \in \mathbb{R}_+^{l_t \times T}$ and given on the path-space by $\bm{\Lambda}^{C,k}_{1:T} = \left(\bm{\Lambda}_1^{C,k},\bm{\Lambda}_2^{C,k}, \ldots,\bm{\Lambda}_T^{C,k}\right)$.
}
\item{Assume that for the random vector $\tilde{\bm{V}}_t^{k}$ for the volume resting in the LOB after the placement of limit orders we have $\tilde{\bm{V}}_t^{k}=\bm{V}_{t-1}^{k}+\bm{N}_t^{LO,k}$, and that the random vector $\bm{N}_t^{C,k} \in \mathbb{N}_+^{l_t}$ is distributed according to a truncated multivariate generalized Cox process with conditional distribution $\bm{N}_t^{C,k}|\tilde{\bm{V}}_t^{k}=\underbar{v} \sim \mathcal{GCP}\left(\bm{\lambda}^{C,k}_t\right)\mathbb{I}(\bm{N}_t^{C,k}<\underbar{v})$ (with $\underbar{v}=(v_{-l_d+1},\ldots,v_{l_p})$) given by
\begin{align} \label{EqnCntsTruncCancel}
\mathbb{P}\text{r}\left(\left. N_t^{C,k,-l_d+1} = n_{-l_d+1},\ldots,N_t^{C,k,l_p} = n_{l_p}
\right|\bm{\Lambda}^{C,k}_t = \bm{\lambda}^{C,k}_t, \tilde{\bm{V}}_t^{k}=\underbar{v} \right) =\prod_{s=-l_d+1}^{l_p} \frac{\frac{(\lambda_t^{C,k,s})^{n_s}}{n_s!}}{\sum_{j=0}^{v_s} \frac{(\lambda_t^{C,k,s})^j}{j!}}.
\end{align}
}
\item{Assume that for the cancellation count $N_t^{C,k,s}$, the orders with highest priority are cancelled from level $s$ (which are also the oldest orders in their respective queue). Assume also that cancellations always remove an order in full, i.e. there are no partial cancellations. 
}
\end{enumerate}
\end{definition}

\begin{remark}
Cancellations are a critical part of a market makers ability to modulate and adjust their liquidity activity to avoid large losses in trades that would otherwise be executed under an adverse selection setting. Under this proposed model for market maker liquidity removal activity (cancellations), the number of limit orders cancelled by the liquidity providers in the market has an appropriate dynamic intensity structure that can evolve intra-daily to reflect the changing nature of liquidity demand throughout the trading day. In addition, the number of limit orders cancelled at each level of the bid and ask also allow for the model to capture the observed dependence structures in order cancellations at each level of the bid and ask. The dependence structure utilised is based on a skew-t copula which allows non-exchangeability of the stochastic intensity on the bid and ask at each level of the book, as well as asymmetry in the tail dependence features. This means that when large price movements occur in the LOB, market makers need to adjust their LOB volumes and profile by cancelling existing resting orders and creating new orders. This typically occurs many times throughout the trading day, and the ability to do this with an appropriate dependence structure is critical. In addition, the number of cancelled orders needs to preserve the principle of volume preservation, that is the upper bound on the total number of Limit Orders that may be cancelled at any given time is based on the instantaneous resting volume in the book at the given time instant.
\end{remark}

We complete the specification of the representative agents by considering the specification of the liquidity demander agent. In addition to market makers who are incentivized to place orders intra-daily in the limit order book, by exchanges in which they operate, there are also other market participants who trade for other reasons. These other market participants include hedge funds, pension funds and other types of large investors, typically we refer to such groups of traders as liquidity demanders. They absorb liquidity throughout the day by purchasing resting orders in the limit order book. These purchases are most often made through market orders or aggressive limit orders. In this chapter we assume that all such activities can be adequately modelled by a stochastic liquidity demander agent making dynamically evolving decisions to place market orders, as specified below.

\begin{definition}[\textbf{Market order submission process for liquidity demander agent}]
Consider a representative agent for the liquidity providers to be composed of a \textbf{market order} component, which has the following stochastic structure:
\begin{enumerate}
\item{Assume a path-space random vector $N_{1:T}^{MO,k}\in \mathbb{N}_+^{1 \times T}$ for the number of market orders constructed from the random variables for the number of market orders in each interval $N_{1:T}^{MO,k} = \left(N_1^{MO,k},N_2^{MO,k}, \ldots,N_T^{MO,k}\right)$. Furthermore, assume that for these random variables the latent stochastic process for the intensity is given by random variable $\Lambda^{MO,k}_{1:T} \in \mathbb{R}_+^{l_t \times T}$, and given on the path-space by $\Lambda^{MO,k}_{1:T} = \left(\Lambda_1^{MO,k},\Lambda_2^{MO,k}, \ldots,\Lambda_T^{MO,k}\right)$.
}
\item{Assume the conditional independence property for the random variables 
\begin{equation}
\left[N_{s}^{MO,k}| \Lambda^{MO,k}_{s}\right] \independent \left[N_{t}^{MO,k}|\Lambda^{MO,k}_{t} \right], \;\; \forall s \neq t,\;\; s,t \in \left\{1,2,\ldots,T\right\}.
\end{equation}
}
\item{Assume that for the random variable $\tilde{R}_t^{k}$ for the volume resting on the opposite side of the LOB after the placement of limit orders and cancellations we have $\tilde{R}_t^{k}=\Sigma_{s=1}^{l_p} \left [ \tilde{V}_{t-\Delta t}^{k',s}-N_t^{C,k',s} \right ]$, where $k'=a$ if $k=b$, and vice-versa, and that the random variable $N_t^{MO,k} \in \mathbb{N}_+$ is distributed according to a truncated generalized Cox process with conditional distribution $N_t^{MO,k}|\tilde{R}_t^{k}=r \sim \mathcal{GCP}\left(\lambda^{MO,k}_t\right)\mathbb{1}(N_t^{MO,k}<r)$ given by
\begin{align} \label{EqnCntsTruncMO}
\mathbb{P}\text{r}\left(\left. N_t^{MO,k} = n \right|\Lambda^{MO,k}_t = \lambda^{MO,k}_t, \tilde{R}_t^{k}=r \right) = \frac{\frac{(\lambda_t^{MO,k})^{n}}{n!}}{\sum_{j=0}^{r} \frac{(\lambda_t^{MO,k})^j}{j!}}.
\end{align}
}
\item{Assume the independence property for random vectors of latent intensities unconditionally according to
\begin{equation}
\Lambda_{s}^{MO,k} \independent \Lambda_{t}^{MO,k}, \;\; \forall s \neq t,\;\; s,t \in \left\{1,2,\ldots,T\right\}.
\end{equation} 
}
\item{Assume that for each intensity random variable $\Lambda^{MO,k}_t \in \mathbb{R}_+$ there is a corresponding transformed intensity variable $\Gamma^{MO,k}_t \in \mathbb{R}$ and the relationship for each element is given by the mapping
\begin{equation} \label{EqnIntenMO}
\Lambda_t^{MO,k} = \mu_0^{MO,k} F\left(\Gamma_t^{MO,k}\right)
\end{equation}
for some baseline intensity parameter $\mu_0^{MO,k} \in \mathbb{R}_+$ and strictly monotonic mapping $F:\mathbb{R} \mapsto [0,1]$.
}
\item{Assume that the random variables $\Gamma^{MO,k}_t \in \mathbb{R}$, characterizing the intensity before transformation of the Generalized Cox-Process, are distributed in interval $[t-1,t)$ according to a univariate skew-t distribution $\Gamma^{MO,k}_t \sim St(m_t^{MO,k},\beta^{MO,k},\nu^{MO,k},\sigma^{MO,k})$. 
}
\item{Assume that for every element $N_t^{MO,k}$ of market order counts, there is a corresponding random vector $\bm{O}_t^{MO,k,s} \in \mathbb{N}_+^{N_t^{MO,k}}$ of order sizes. We assume that the element $O_{i,t}^{MO,k}, i \in \left \{ 1,\ldots, N_t^{MO,k}\right \} $ is distributed according to $O_{i,t}^{MO,k} \sim H(\cdot)$. Assume also that market order sizes are unconditionally independent $ O_{i,t}^{MO,k} \independent  O_{i',t}^{MO,k}$ for $i \neq i'$ or $t \neq t'$.
}
\end{enumerate}
\end{definition}

We denote the LOB state for the real dataset at time $t$ on a given day by the random vector $\bm{L}_{t}$, and this corresponds to the prices and volumes at each level of the bid and ask. Utilising the stochastic agent-based model specification described above, and given a parameter vector $\bm{\theta}$, which will generically represent all parameters of the liquidity providing and liquidity demanding agent types, one can then also generate simulations of intra-day LOB activity and arrive at the synthetic state $\bm{L}^{\ast}_{t}\left(\bm{\theta}\right)$. The state of the simulated LOB at time $t$ is obtained from the state at time $t-1$ and a set of stochastic components, denoted generically by $\bm{X}_t$, which are obtained from a single stochastic realisation of the following components of the agent-based models:

\begin{itemize}
\item{Limit order submission intensities $\bm{\Lambda}^{LO,b}_t$, $\bm{\Lambda}^{LO,a}_t$, order numbers $\bm{N}^{LO,b}_t$, $\bm{N}^{LO,a}_t$, and order sizes $\bm{O}_{i,t}^{LO,a,s},\allowbreak \bm{O}_{j,t}^{LO,b,s}$, where {$s=-l_d+1 \ldots l_p,i=1 \ldots N_t^{LO,a,s},j=1 \ldots N_t^{LO,b,s}$} };
\item{Limit order cancellation intensities $\bm{\Lambda}^{C,b}_t$, $\bm{\Lambda}^{C,a}_t$ and numbers of cancellations $\bm{N}^{C,b}_t$, $\bm{N}^{C,a}_t$ };
\item{Market order intensities $\bm{\Lambda}^{MO,b}_t$, $\bm{\Lambda}^{MO,a}_t$, numbers of market orders $\bm{N}^{MO,b}_t$, $\bm{N}^{MO,a}_t$,$\bm{V}^{MO,b}_t$,$\bm{V}^{MO,a}_t$ and market order sizes $\bm{O}_{i,t}^{MO,a},\bm{O}_{j,t}^{MO,b},i=1 \ldots N_t^{MO,a},j=1 \ldots N_t^{MO,b}$}
\end{itemize}
These stochastic features are combined with the previous state of the LOB, $\bm{L}^{\ast}_{t-1}\left(\bm{\theta}\right)$, to produce the new state $L^{\ast}_{t}\left(\bm{\theta}\right)$ for a given set of parameters $\bm{\theta}$, given by
\begin{equation}
\bm{L}^{\ast}_{t}\left(\bm{\theta}\right)=G(\bm{L}^{\ast}_{t-1}\left(\bm{\theta}\right),\bm{X}_t)
\end{equation}

$G(\cdot)$ is a transformation that maps the previous state of the LOB and the activity generated in the current step into a new step, much the same way as the matching engine updates the LOB after every event. As we model the activity in discrete intervals, however, the LOB is only updated at the end of every interval, and the incoming events (limit orders, market orders and cancellations) are processed in the order specified in Section \ref{sec:lobmodel}. Conditional then on a realization of these parameters $\boldsymbol{\theta}$, the trading activity in the LOB can be simulated for a single trading day, and the complete procedure is described in the algorithm set out in \cite{panayi2015stochastic}. 

\subsection{Bayesian model formulation of the stochastic agent LOB model representation}
In this section we consider the class of LOB stochastic models developed in the previous section and we detail a Bayesian model formulation under an ABC framework. Methods for Bayesian modelling in the presence of computationally intractable likelihood functions are of growing interest. These methods may arise either because the likelihood is truly intractable to evaluate point-wise, or in our case it may be that the likelihood is so complex in terms of model specification and costly to evaluate point-wise that one has to resort to alternative methods to perform estimation and inference. Termed {\it likelihood-free samplers} or {\it Approximate Bayesian Computation} (ABC) methods, simulation algorithms such as Sequential Monte Carlo Samplers have been adapted for this setting, see for instance \cite{peters2012sequential}. 

We start by recalling a few basics. Typically, Bayesian inference proceeds via the posterior distribution, generically denoted by $\pi(\bm{\theta}|\bm{y})\propto f(\bm{y}|\bm{\theta})\pi(\bm{\theta})$, the updating of prior information $\pi(\bm{\theta})$ for a parameter $\bm{\theta}\in E$ through the likelihood function $f(\bm{y}|\bm{\theta})$ after observing data $\bm{y}\in{\mathcal Y}$.  Numerical algorithms, such as importance sampling, Markov chain Monte Carlo (MCMC) and sequential Monte Carlo (SMC),  are commonly employed to draw samples from the posterior $\pi(\bm{\theta}|\bm{y})$. 

\begin{remark}[Note on Data vector]
In the context of this chapter the data $\bm{y}$ is the entire order book structure for a given asset over a day as summarized by: 
\begin{itemize}
\item{Limit order submission order numbers $\bm{N}^{LO,b}_t$, $\bm{N}^{LO,a}_t$, and order sizes $\bm{O}_{i,t}^{LO,a,s},\allowbreak \bm{O}_{j,t}^{LO,b,s}$, where {$s=-l_d+1 \ldots l_p,i=1 \ldots N_t^{LO,a,s},j=1 \ldots N_t^{LO,b,s}$} }
\item{Limit order numbers of cancellations $\bm{N}^{C,b}_t$, $\bm{N}^{C,a}_t$ }
\item{Numbers of market orders $\bm{N}^{MO,b}_t$, $\bm{N}^{MO,a}_t$,$\bm{V}^{MO,b}_t$,$\bm{V}^{MO,a}_t$ and market order sizes \\ $\bm{O}_{i,t}^{MO,a},\bm{O}_{j,t}^{MO,b},i=1 \ldots N_t^{MO,a},j=1 \ldots N_t^{MO,b}$.}
\end{itemize}
The resulting observation vector $\bm{y}_t$, at time $t$, is then the concatenation of all these variables. These stochastic features are obtained at sampling rate $t$ within the market hours of the trading day, typically say every 5-30 seconds for the 8.5 hour trading day, producing a total of between 1000-6000 vector valued observations per day. 
\end{remark}

Clearly, even evaluating a likelihood on this many records, even if it could be written down which in many LOB models built on queues like the one in this chapter will not be the case, would still be a challenging task. Generically we will denote below the collection of all observations $\bm{y}$ of the LOB for an asset on a given day, and by $\bm{\theta}$ the set of all parameters that are utilised to parameterize the LOB stochastic model.

There is growing interest in posterior simulation in situations  where the likelihood function is computationally intractable i.e.  $f(\bm{y}|\bm{\theta})$ may not be  numerically evaluated pointwise. As a result, sampling algorithms based on repeated likelihood evaluations require modification for this task. 

Collectively known as {\it likelihood-free samplers} (and also as {\it approximate Bayesian computation}) these methods have been developed across multiple disciplines. They employ generation of auxiliary datasets under the model as a means to circumvent (intractable) likelihood evaluation.

\subsubsection{Posterior models for computationally intractable likelihoods}
\label{section:abc}

In essence, likelihood-free methods first reduce the observed data, $\bm{y}$,  to a low-dimensional vector of summary statistics $t_{y}=T(\bm{y})\in{\mathcal T}$, where $\dim(\bm{\theta})\leq\dim(t_y)<<\dim(\bm{y})$. Accordingly, the true posterior $\pi(\bm{\theta}|\bm{y})$ is replaced with a new posterior $\pi(\bm{\theta}|t_y)$. These are equivalent if $t_y$ is sufficient for $\bm{\theta}$, and $\pi(\bm{\theta}|t_y)\approx\pi(\bm{\theta}|\bm{y})$ is an approximation if there is some loss of information through $t_y$.
The new target posterior, $\pi(\bm{\theta}|t_y)$, still assumed to be computationally intractable, is then
embedded within an augmented model from which sampling  is viable. Specifically the joint posterior of the
model parameters $\bm{\theta}$, and auxiliary data $t\in{\mathcal T}$ given observed data $t_y$ 
is 
\begin{equation}
\label{eqn:joint}
	\pi(\bm{\theta},t|t_y)\propto K_h(t_y-t)f(t|\theta)\pi(\theta),
\end{equation}
where 
$t\sim f(t|\bm{\theta})$ may be interpreted as the vector of summary statistics $t=T(x)$ computed from a dataset simulated according to the model $\bm{x}\sim f(\bm{x}|\bm{\theta})$. Assuming such simulation is possible, data-generation under the model, $t\sim f(t|\bm{\theta})$, forms the basis of computation in the likelihood-free setting. The target marginal posterior $\pi_M(\bm{\theta}|t_y)$ for the parameters $\bm{\theta}$, is then obtained as 
\begin{equation}
\label{eqn:marginal}
	\pi_M(\bm{\theta}|t_y) = c_M\int_{\mathcal{T}}K_h(t_y-t)f(t|\bm{\theta})\pi(\bm{\theta}) dt
\end{equation}
where $(c_M)^{-1}=\int_{E}\int_{\mathcal{T}}K_h(t_y-t)f(t|\bm{\theta})\pi(\bm{\theta}) dtd\bm{\theta}$ normalises (\ref{eqn:marginal}) such that it is a density in $\bm{\theta}$ (e.g. \cite{reeves2005theoretical}; \cite{wilkinson2013approximate}; \cite{blum2010approximate}; \cite{sisson2007sequential}; \cite{fearnhead2012constructing}).
The function $K_h(t_y-t)$ is a standard kernel function, with scale parameter $h\geq 0$, which weights the intractable posterior with high density in regions $t\approx t_y$ where auxiliary and observed datasets are similar. As such, $\pi_M(\bm{\theta}|t_y)\approx\pi(\bm{\theta}|t_y)$ forms an approximation to the intractable posterior via (\ref{eqn:marginal}) through standard smoothing arguments (e.g. \cite{blum2010approximate}).
In the  case as $h\rightarrow 0$, so that $K_h(t_y-t)$ becomes a point mass at the origin (i.e. $t_y=t$) and is zero elsewhere,  if $t_y$ is sufficient for $\theta$ then the intractable posterior marginal $\pi_M(\bm{\theta}|t_y)=\pi(\bm{\theta}|t_y)=\pi(\bm{\theta}|\bm{y})$ is recovered exactly (although small $h$ is usually impractical). Various choices of smoothing kernel $K$ have been examined  in the literature (e.g. \cite{marjoram2003markov}; \cite{beaumont2002approximate}; \cite{peters2012sequential}).

For our discussion on likelihood-free or ABC samplers, it is convenient to consider a generalisation of the joint distribution  (\ref{eqn:joint}) incorporating $S\geq 1$ auxiliary summary vectors
\begin{equation*}
	\pi_J(\bm{\theta},t^{1:S}|t_y)
	 \propto
	\tilde{K}_h(t_y,t^{1:S})f(t^{1:S}|\bm{\theta})\pi(\bm{\theta})
\end{equation*}
where $t^{1:S}=(t^1,\ldots,t^S)$  and $t^1,\ldots,t^S\sim f(t|\bm{\theta})$ are $S$ independent datasets generated from the (intractable) model. As the auxiliary datasets are, by construction, conditionally independent given $\bm{\theta}$, we have $f(t^{1:S}|\bm{\theta})=\prod_{s=1}^Sf(t^s|\bm{\theta})$.
We follow \cite{del2012adaptive} and specify the kernel $\tilde{K}$  as $\tilde{K}_h(t_y,t^{1:S})=S^{-1}\sum_{s=1}^SK_h(t_y-t^s)$,
which produces the joint posterior
\begin{equation}
\label{eqn:jointS}
	\pi_J(\bm{\theta},t^{1:S}|t_y)
	 =
	 c_J
	\left[\frac{1}{S}\sum_{s=1}^SK_h(t_y-t^s)\right]
	\left[\prod_{s=1}^Sf(t^s|\bm{\theta})\right]
	\pi(\bm{\theta}),
\end{equation}
with $c_J>0$ the appropriate normalisation constant, where in (\ref{eqn:jointS}) we extend the uniform kernel choice of $K(t_y-t^s)$ by \cite{del2012adaptive} to the general case. It is easy to see that, by construction,  $\int_{{\mathcal T}^S}\pi_J(\bm{\theta},t^{1:S}|t_y)dt^{1:S}=\pi_M(\bm{\theta}|t_y)$ admits the distribution (\ref{eqn:marginal}) as a marginal distribution (c.f. \cite{del2012adaptive}).
The case  $S=1$ with $\pi_J(\theta,t^{1:S}|t_y)=\pi(\theta,t|t_y)$ corresponds to the more usual joint posterior (\ref{eqn:joint}) in the likelihood-free setting.

There are two obvious approaches to posterior simulation from $\pi_M(\bm{\theta}|t_y)\approx\pi(\bm{\theta}|t_y)$ as an approximation to $\pi(\bm{\theta}|\bm{y})$. The first approach proceeds by sampling directly on the augmented model $\pi_J(\bm{\theta},t^{1:S}|t_y)$, realising joint samples $(\bm{\theta},t^{1:S})\in E\times{\mathcal T}^S$ before {\it a posteriori} marginalisation over $t^{1:S}$ (i.e. by discarding the $t^s$ realisations from the sampler output). In this approach, the summary quantities $t^{1:S}$ are treated as parameters in the augmented model.
 
The second approach is to sample from $\pi_M(\bm{\theta}|t_y)$ directly, a lower dimensional space, by approximating the integral (\ref{eqn:marginal}) via Monte Carlo integration in lieu of each posterior evaluation of $\pi_M(\bm{\theta}|t_y)$. In this case
\begin{equation}
\label{eqn:abc-monte-carlo}
	\pi_M(\bm{\theta}|t_y) 
	\propto 
	\pi(\bm{\theta})\int_{\mathcal T}K_h(t_y-t)f(t|\bm{\theta})dt
	 \approx 
	\frac{\pi(\bm{\theta})}{S}\sum_{s=1}^S K_h(t_y-t^{s})
	:=\hat{\pi}_M(\bm{\theta} | t_y),
\end{equation}
where $t^1,\ldots,t^S\sim f(t|\bm{\theta})$. This expression,  examined by various  authors (e.g. \cite{marjoram2003markov}; \cite{reeves2005theoretical}; \cite{ratmann2009model}; \cite{toni2009approximate}; \cite{peters2012sequential}), requires multiple generated datasets $t^1,\ldots,t^ S$, for each evaluation of the marginal posterior distribution $\pi_{M}(\bm{\theta}|t_y)$. As with standard Monte Carlo approximations, $\mbox{Var}[\hat{\pi}_M(\bm{\theta}|t_y)]$ reduces as $S$ increases, with $\lim_{S\rightarrow\infty}\mbox{Var}[\hat{\pi}_M(\bm{\theta}|t_y)]=0$. For the marginal posterior distribution, the quantities $t^{1:S}$ serve only as a means to estimate $\pi_M(\bm{\theta}|t_y) $, and do not otherwise enter the model explicitly. The number of samples $S$ directly impacts on the variance of the estimation.

\section{Estimation of Bayesian LOB stochastic agent model via Population-based samplers}
\label{sec:estimation}

Population-based likelihood-free samplers  were introduced to circumvent poor mixing in  MCMC samplers (\cite{sisson2007sequential};  \cite{toni2009approximate}; \cite{beaumont2009adaptive}; \cite{peters2012sequential}; \cite{del2012adaptive}). These samplers propagate a population of {\it particles}, $\bm{\theta}^{(1)},\ldots,\bm{\theta}^{(N)}$, with associated importance weights $W(\bm{\theta}^{(i)})$, through a sequence of related densities $\phi_1(\bm{\theta}_1),\ldots,\phi_{T}(\bm{\theta}_T)$, which defines a smooth transition from the distribution $\phi_1$, from which direct sampling is available, to $\phi_T$ the target distribution. 

For likelihood-free or ABC samplers, $\phi_k$ is defined by allowing $K_{h_n}(t_y-t)$ to place greater density on regions for which $t_y\approx t$ as $k$ increases (that is, the bandwidth $h_n$ decreases with $n$). Hence,  we denote $\pi_{J,n}(\bm{\theta}, t^{1:S}|t_y)\propto \tilde{K}_{h_n}(t_y,t^{1:S})f(t^{1:S}|\bm{\theta})\pi(\bm{\theta})$ and $\pi_{M,n}(\bm{\theta}|t_y)\propto\pi(\bm{\theta})\int_{\mathcal{T}^S}\tilde{K}_{h_n}(t_y,t^{1:S})f(t^{1:S}|\bm{\theta})dt^{1:S}$ for $n=1,\ldots,T$, under the joint and marginal posterior models respectively.

\subsection{Sequential Monte Carlo-based samplers}

SMC  methods have  emerged out of the fields of engineering, probability and statistics in recent years.
Variants of the methods sometimes appear under the names of particle filtering or interacting  particle systems (e.g.,\cite{ristic2004beyond},\cite{doucet2001introduction}, \cite{del2004feynman}), and their theoretical properties have been extensively studied  by \cite{crisan2002survey}, \cite{del2004feynman}, \cite{chopin2004central}, and \cite{kunsch2005recursive}. 

The standard SMC algorithm involves finding a numerical solution to a set of filtering recursions, such as filtering problems arising from nonlinear/non-Gaussian state space models.  Under this framework, the SMC algorithm samples from an (often naturally occurring) sequence of distributions $\pi_{n}$, indexed by $n=1,\ldots,T$.   Each distribution is defined on the support $E^{n}=E\times E\times \cdots \times E$ for some generic space denoted $E$. Recently, this class of algorithms was adapted to tackle the same class of problems typically addressed by MCMC methods where one has instead a sequence of distributions $\left\{\pi_n\right\}_{n \geq 1}$ each defined on fixed support $E$. NOTE: this is not a product space but a fixed space $E$. \cite{del2006sequential}, \cite{peters2005topics}, \cite{peters2012sequential} and \cite{targino2015sequential} generalize the SMC algorithm to the case where the target distributions $\pi_n$ are all defined on the  same support $E$. This generalization, termed the SMC {\it sampler},  adapts the SMC algorithm to the more popular setting in which the state space $E$ remains static, that is, the settings we have discussed earlier with regard to the MCMC algorithms.

In short, the SMC sampler generates weighted samples (termed {\it particles}) from a sequence of
distributions $\pi_n$, for $n=1,\ldots, T$, where $\pi_T$ may be of particular interest. We refer to $\pi_T$ as the target distribution such as a posterior distribution for model parameters. Procedurally,  particles obtained from an arbitrary initial distribution $\pi_1$, with a set of corresponding initial weights,  are sequentially propagated through each distribution  $\pi_t$ in the sequence via three processes, involving mutation (or move), correction (or importance weighting), and selection (or resampling). The final weighted particles at distribution $\pi_T$ are \hbox{considered} weighted samples from the target distribution $\pi$. 

Hence, given a sequence of distributions $\left\{ \pi_n(d\bm{\theta}) \right\}_{n=1}^T$, the aim is to develop a large collection of $N$-weighted random samples at each time $n$ denoted by $\left\{W_n^{(i)},\bm{\Theta}_n^{(i)}\right\}_{i=1}^N$ such that $W_n^{(i)}\,{>}\,0$ and $\sum_{i=1}^N W_n^{(i)} \,{=}\, 1$. These importance weights and samples, denoted by $\left\{W_n^{(i)},\bm{\Theta}_n^{(i)}\right\}_{i=1}^N$, are known as particles (hence the name often given to such algorithms as particle filters or interacting particle systems). For such approaches to be sensible we would require that the empirical distributions constructed through these samples should converge asymptotically $(N\,{\rightarrow}\, \infty)$ to the target distribution $\pi_n$ for each time $n$. This means that for any $\pi_n$ integrable function, denoted, for example,  by $\phi(\bm{\theta}): E \rightarrow \mathbb{R}^{\prime}$ one would have the following convergence:
\begin{align}
\sum_{i=1}^N W_n^{(i)}\phi\left( \bm{\theta}_{n}^{(i)}\right) \stackrel{a.s.}{\rightarrow} \mathbb{E}_{\pi_n}\big[\phi(\bm{\Theta})\big].
\end{align}

In the SMC Sampler algorithm, a particular variant of SMC algorithms, a modification of the SMC algorithm, is developed. Consider a generic sequence of distributions given by $\pi_n(\bm{\theta}),  n=1,\ldots, T$, with $\bm{\theta} \in E$, where the final distribution $\pi_T$ is the distribution of interest. By introducing a sequence of backward kernels $L_k$,  a new distribution
\begin{align} \label{SMCSamplEqns}
\widetilde{ \pi }_{n} (\bm{\theta}_1, \ldots, \bm{\theta}_n ) =\pi _{n} ( \bm{\theta}_{n}) \prod\limits_{k=1}^{n-1}L_{k}\left( \bm{\theta}_{k+1},\bm{\theta}_{k}\right)
\end{align}
may be defined for the {\it path} of a particle $(\bm{\theta}_1, \ldots, \bm{\theta}_n) \in E^n$ through the sequence $\pi_1,\ldots,\pi_n$.  The only restriction on the backward kernels is that the correct marginal distributions \\ $\int \widetilde{\pi} _{n}( \bm{\theta}_1, \ldots, \bm{\theta}_n ) d\bm{\theta}_1, \ldots, d\bm{\theta}_{n-1} \,{=}\, \pi_n(\bm{\theta}_n)$  are available. Within this framework, one may then work with the constructed sequence of distributions, $\widetilde{ \pi }_n$, under the standard SMC algorithm.

In summary, the SMC Sampler algorithm involves three stages:
\begin{enumerate}
\item{{\it Mutation}, whereby the particles are moved from $\bm{\theta}_{n-1}$ to $\bm{\theta}_n$ via a mutation kernel $M_n(\bm{\theta}_{n-1},\bm{\theta}_n)$;}
\item{{\it Correction},  where the particles are reweighted with respect to $\pi_n$ via the incremental importance weight (Eq. \ref{Eqnweight});}
\item{ {\it Selection}, where according to some measure of particle diversity, commonly the effective sample size, the weighted particles may be resampled in order to reduce the variability of the importance weights. }
\end{enumerate}

In more detail, suppose that at time $n-1$, the distribution $\widetilde{ \pi }_{n-1}$ can be approximated empirically by $\widetilde{ \pi }_{n-1}^{N}$ using $N$-weighted particles. These particles are first propagated to the next distribution $\widetilde{ \pi }_{n}$ using a mutation kernel $M_n(\bm{\theta}_{n-1},\bm{\theta}_n)$, and then assigned new weights $W_n = W_{n-1}w_{n}\left( \bm{\theta}_1, \ldots \bm{\theta}_n\right)$, where $W_{n-1}$ is the weight of a particle at time $n-1$ and $w_{n}$ is the incremental importance weight given by
\begin{align}\label{Eqnweight}
w_{n}\left( \bm{\theta}_1,\ldots, \bm{\theta}_n\right) =\frac{\widetilde{\pi }_{n}\left(
\bm{\theta}_1,\ldots, \bm{\theta}_n\right) }{\widetilde{\pi }_{n-1}\left( \bm{\theta}_1, \ldots, \bm{\theta}_{n-1}\right)
M_{n}\left( \bm{\theta}_{n-1},\bm{\theta}_{n}\right) }=\frac{\pi _{n}\left(
\bm{\theta}_{n}\right) L_{n-1}\left( \bm{\theta}_{n},\bm{\theta}_{n-1}\right) }{\pi _{n-1}\left(
\bm{\theta}_{n-1}\right) M_{n}\left( \bm{\theta}_{n-1},\bm{\theta}_{n}\right) }.
\end{align}

The resulting particles are now weighted samples from   $\widetilde{ \pi }_{n}$. Consequently, from Eq. (\ref{Eqnweight}), under the SMC Sampler framework, one may work directly with the marginal \nobreak{distributions} $\pi_n(\bm{\theta}_n)$ such that $w_n(\bm{\theta}_1,\ldots,\bm{\theta}_n)=w_n(\bm{\theta}_{n-1},\bm{\theta}_n)$. While the choice of the backward kernels $L_{n-1}$ is essentially arbitrary,  their specification can strongly affect the performance of the algorithm, as will be discussed in the following subsections. The basic version of the SMC Sampler algorithm therefore proceeds explicitly as given in Algorithm \ref{Chapter_SMCSampler_algorithm}.

\begin{remark} In all cases in which we utilize the incremental importance sampling weight correction, the arguments in the expressions only need to be known up to normalization. That is, it is perfectly acceptable to only be able to evaluate the sequence of target distributions $\left\{\pi_n\right\}$ up to normalization constant. This is true as long as the same normalization constant is present for all particles, since the renormalization step will correct for this lack of knowledge in the importance weighting. In practice, this is critical to the application of such methods.
\end{remark}

\textbf{Sequential Monte Carlo Samplers} \label{Chapter_SMCSampler_algorithm}
\begin{enumerate}
\item{Initialize the particle system;}
\begin{enumerate}
\item{Set $n = 1$;}
\item{ For $i = 1,\ldots,N$, draw initial particles $\bm{\Theta}_{1}^{(i)} \sim p(\bm{\theta})$;}
\item{Evaluate incremental importance weights $\left\{w_1\left(\bm{\Theta}_1^{(i)}\right)\right\}$ using Equation (\ref{Eqnweight}) and normalize the weights to obtain $\left\{W_1^{(i)}\right\}$.}
\end{enumerate}
\item[]{Iterate the following steps through each distribution in sequence $\left\{\pi_t\right\}_{n=2}^T$.}
\item{Resampling}
\begin{enumerate}
\item{If the effective sampling size $(ESS) = \frac{1}{\sum_{i=1}^N \left(w^{(i)}_n\right)^2} < N_{eff}$ is less than a threshold $N_{eff}$, then resample the particles via the empirical distribution of the weighted sample either by multinomial or stratified methods; see discussions on unbiased resampling schemes by \cite{kunsch2005recursive} and \cite{del2004feynman}.}
\end{enumerate}
\item{Mutation and correction}
\begin{enumerate}
\item{Set $n=n+1$, if $n = T+1$, then stop;}
\item{For $i = 1,\ldots,N$ draw samples from mutation kernel $\bm{\Theta}_n^{(i)} \sim M_{n}\left(\bm{\Theta}_{n-1}^{(i)}\right)$;}
\item{Evaluate incremental importance weights $\left\{w_n\left(\bm{\Theta}_n^{(i)}\right)\right\}$ using Equation (\ref{Eqnweight}) and\break normalize the weights to obtain $\left\{W_n^{(i)}\right\}$ via
\begin{align}
W_{n}^{(i)} = W_{n-1}^{(i)} \frac{w^{(i)}_n\left(\bm{\Theta}_{n-1},\bm{\Theta}_{n}\right)}{\sum_{j=1}^N W_{n-1}^{(i)} w^{(i)}_n\left(\bm{\Theta}_{n-1},\bm{\Theta}_{n}\right)}.
\end{align}
}
\end{enumerate}
\end{enumerate}

At this stage, for practitioners wishing to utilise such SMC methods, it is informative to understand better what properties are known about this class of estimation methods in terms of accuracy of such numerical approximations. We mention briefly some known results based on recent examples of concentration inequalities for particle methods that are finite sample result (see discussion and references by \cite{del2013introduction}).

The exponential concentration inequalities presented here are satisfied under some regularity conditions on the particle weights and the mutation kernel $M_n$ when defined on some general state space $E_n$; see specific probabilistic details of these conditions by \cite{del2004feynman}.

Using the concentration analysis of mean field particle models, the following exponential estimate can be obtained (see discussion by \cite{del2004feynman}) and references therein. Note in the following when the $N$ particle approximation to a distribution or density, such as $\pi$, is used we will denote it by $\pi^N$.

\begin{theorem}[Finite Sample Exponential Concentration Inequality]
For any $x\geq 0$, ${n\geq 0}$, and any population size $N\geq 1$, the probability of the event is\vspace*{-3pt}
\begin{align}
\mathbb{P}\mathrm{r}\left(\left|\pi^N_{n}(\varphi) - \pi_{n}(\varphi)\right| \leq \frac{c_1}{N}~\left(1+x+\sqrt{x}\right)+\frac{c_2}{\sqrt{N}}~\sqrt{x}\right) \geq 1-e^{-x},
\end{align}
where one defines the $N$ particle sample estimator as follows:\vspace*{-3pt}
\begin{align*}
\pi^N_{n}(\varphi) = \sum_{i=1}^N W_n^{(i)}\varphi\left(\bm{\theta}_n^{(i)}\right)
\end{align*}
and\vspace*{-3pt}
\begin{align}
\pi_{n}(\varphi) = \int \varphi\left(\bm{\theta}_n\right)\pi_n\left(\bm{\theta}_n\right)~ d\bm{\theta}_n.
\end{align}
\end{theorem}

In the case of a stable SMC algorithm, that is, one that is insensitive to initial conditions, such as those we discussed earlier, the constants $c$ and $(c_1,c_2)$ do not depend on the time parameter. One can also bound the difference between the particle estimate of the target distribution and the true distribution as follows. Consider that for any $\bm{\theta}=(\theta_i)_{1\leq i\leq d}$ and any $(-\infty,x]=\prod_{i=1}^d(-\infty,\theta_i]$ cells in $E_n=\mathbb{R}^d$, we let\vspace*{-3pt}
\begin{align*}
F_n(x)=\pi_n\left(\mathbb{1}_{(-\infty,x]}\right)\quad\mbox{\rm and}\quad
F^N_n(x)=\pi^N_n\left(\mathbb{1}_{(-\infty,x]}\right).
\end{align*}
Using these definitions of the empirical particle constructed distribution function and the target distribution function at sequence number $n$ in the sequence of distribution $\left\{\pi_1,\pi_2,\ldots,\pi_T\right\}$, we can state the following corollary for the distribution functions for sequence of densities $\pi_t$ given previously.

\begin{corollary}
For any $y\geq 0$, $n\geq 0$, and any population size $N\geq 1$, the probability of the following event\vspace*{-3pt}
 \begin{align*}
\sqrt{N}~\left\|F^N_n-F_n\right\|\leq c~\sqrt{d~(y+1)}
\end{align*}
is greater than $1-e^{-y}$.
\end{corollary}

This concentration inequality ensures that the particle repartition function $F^N_t$ converges to $F_t$, almost surely for the uniform norm. 

\subsection{Sequential Monte Carlo samplers for intractable likelihood Bayesian models}
Formalizing this in the context of SMC Samplers for ABC posterior settings, the particle population $\bm{\theta}_{n-1}$ drawn from the distribution $\phi_{n-1}(\bm{\theta}_{n-1})$ at time $n-1$ is mutated to $\phi_n(\bm{\theta}_n)$ by the kernel $M_n(\bm{\theta}_{n-1},\bm{\theta}_n)$. The weights  for the mutated particles $\theta_n$ may be obtained as
$
	W_n(\bm{\theta}_n) = W_{n-1}(\bm{\theta}_{n-1})w_{n}\left(\bm{\theta}_{n-1}, \bm{\theta}_n\right)
$
where, for the marginal model sequence $\pi_{M,n}(\bm{\theta}_n|t_y)$, the incremental weight is 
\begin{equation}
\label{eqn:smc-weight-increment}
	w_{n}\left( \bm{\theta}_{n-1}, \bm{\theta}_n\right) 
	= 
	\frac{\pi_{M, n}(\bm{\theta}_{n}|t_y) L_{n-1}\left( \bm{\theta}_{n},\bm{\theta}_{n-1}\right) }{\pi _{M, n-1}(\bm{\theta}_{n-1}|t_y) M_{n}\left( \bm{\theta}_{n-1},\bm{\theta}_{n}\right) }
	\approx
	\frac{\hat{\pi}_{M, n}(\bm{\theta}_{n}|t_y) L_{n-1}\left( \bm{\theta}_{n},\bm{\theta}_{n-1}\right) }{\hat{\pi} _{M, n-1}(\bm{\theta}_{n-1}|t_y) M_{n}\left( \bm{\theta}_{n-1},\bm{\theta}_{n}\right) },
\end{equation}
where, following (\ref{eqn:abc-monte-carlo}), and setting the kernel bandwidth to an ABC tolerance level $\epsilon_n$ we obtain 
\begin{equation*}
	\hat{\pi}_{M, n}(\bm{\theta}_{n}|t_y):=\frac{\pi(\bm{\theta})}{S}\sum_{s=1}^S K_{\epsilon_n}(t_y-t^{s})
\end{equation*}
which is proportional to an (unbiased) estimate of $\pi_{M, n}(\bm{\theta}_{n}|t_y)$ based on $S$ Monte Carlo draws $t^1,\ldots,t^S\sim f(t|\bm{\theta}_n)$. Here $L_{n-1}\left( \bm{\theta}_{n},\bm{\theta}_{n-1}\right)$ is a reverse-time kernel describing the mutation of particles from $\phi_n(\bm{\theta}_n)$ at time $n$ to $\phi_{n-1}(\bm{\theta}_{n-1})$ at time $n-1$. As with the ABC-MCMC algorithm, the incremental weight (\ref{eqn:smc-weight-increment}) consists of the ``biased'' ratio $\hat{\pi}_{M,n}(\bm{\theta}_n|t_y)/\hat{\pi}_{n-1}(\bm{\theta}_{M,n-1}|t_y)$ for finite $S\geq1$.

If we now consider a sequential Monte Carlo sampler under the joint model $\pi_{J, n}(\bm{\theta}, t^{1:S}|t_y)$, with the natural mutation kernel factorisation
\[
	M_n[(\bm{\theta}_{n-1},t_{n-1}^{1:S}),(\bm{\theta}_{n},t_n^{1:S})]=M_n(\bm{\theta}_{n-1},\bm{\theta}_n)\prod_{s=1}^Sf(t_n^{s}|t_y)
\]
(and similarly for $L_{n-1}$),  following the form of (\ref{eqn:smc-weight-increment}), the incremental weight is exactly
\begin{equation}
\label{eqn:smc-weight-increment-J}
	w_{n}\left[ (\bm{\theta}_{n-1}, t_{n-1}^{1:S}), (\bm{\theta}_n, t_n^{1:S})\right] 
	= 
	\frac{\frac{1}{S}\sum_sK_{\epsilon_n}(t_y-t_n^{s})\pi(\bm{\theta}_n) L_{n-1}\left( \bm{\theta}_{n},\bm{\theta}_{n-1}\right)}{\frac{1}{S}\sum_sK_{\epsilon_{n-1}}(t_y-t_{n-1}^{s})\pi(\bm{\theta}_{n-1}) M_{n}\left( \bm{\theta}_{n-1},\bm{\theta}_{n}\right)}.
\end{equation}

Hence, as the incremental weights (\ref{eqn:smc-weight-increment}, \ref{eqn:smc-weight-increment-J}) are equivalent, they induce identical SMC algorithms for both marginal and joint models $\pi_{M}(\bm{\theta}|t_y)$ and $\pi_{J}(\bm{\theta},t^{1:S}|t_y)$. As a result, while applications of the marginal sampler targeting $\pi_{M}(\bm{\theta}|y)$ are theoretically biased for finite $S\geq 1$, as before, they are in practice unbiased through association with the equivalent sampler on joint space targeting $\pi_{J}(\bm{\theta},t^{1:S}|t_y)$.

We note that a theoretically unbiased sampler targeting $\pi_M(\bm{\theta}|t_y)$, for all $S\geq 1$, can be obtained by careful choice of the kernel $L_{n-1}(\bm{\theta}_n,\bm{\theta}_{n-1})$. 
For example, \cite{peters2005topics}, \cite{peters2012sequential} and \cite{targino2015sequential} all use the suboptimal approximate optimal kernel given by
\begin{equation}
\label{eqn:L3}
    L_{n-1}(\bm{\theta}_n, \bm{\theta}_{n-1})=\frac{\pi_{M,n-1}(\bm{\theta}_{n-1}|t_y)M_n(\bm{\theta}_{n-1},
    \bm{\theta}_n)}
        {\int\pi_{M, n-1}(\bm{\theta}_{n-1}|t_y)M_n(\bm{\theta}_{n-1},\bm{\theta}_n)d\bm{\theta}_{n-1}},
\end{equation}
from which the incremental weight (\ref{eqn:smc-weight-increment}) is approximated by
\begin{eqnarray}
\label{eqn:L3app}
    w_n(\bm{\theta}_{n-1},\bm{\theta}_n)
    & = &
    \pi_{M,n}(\bm{\theta}_n|t_y)/\int \pi_{M, n-1}(\bm{\theta}_{n-1}|t_y)M_n(\bm{\theta}_{n-1},\bm{\theta}_n)d\bm{\theta}_{n-1}\nonumber\\
    & \approx & 
    \hat{\pi}_{M, n}(\bm{\theta}_n|t_y)/\sum_{i=1}^N W_{n-1}(\bm{\theta}^{(i)}_{n-1})M_n(\bm{\theta}^{(i)}_{n-1},\bm{\theta}_n) .
\end{eqnarray}
Under this choice of backward kernel, the weight calculation  is now unbiased for all $S \geq 1$, since the approximation $\hat{\pi}_{M,n-1}(\bm{\theta}|y)$ in the denominator of (\ref{eqn:smc-weight-increment}) is no longer needed. 

\subsubsection{Adaptive Schedules: Choice of sequence of ABC Bayesian distributions via annealed tolerance schedule}
\label{sec:tolschedule}
In this section we consider how to take the ABC posterior distribution and construct a the sequence of distributions that are required for the SMC Sampler. That is, we address the question of how to develop an ABC specific sequence of target distributions. We have chosen to design this sequence by following what we call ABC reverse annealing. In particular, we construct a sequence of target posterior distributions $\left\{\phi_n\right\}_{n \geq 0}$ which are constructed based on strictly decreasing tolerance values, generically denoted by the sequence $\left\{\epsilon_n\right\}_{n \geq 0}$ such that $\epsilon_1 > \epsilon_2 > \ldots > \epsilon_n > \ldots >\epsilon_T$. We obtain this sequence of ABC posterior distributions by considering the $\phi_n$, which was defined with respect to the ABC likelihood involving a kernel. If we consider the kernel to have a decreasing bandwidth given by $K_{\epsilon_n}(t_y-t)$ then we will progressively place greater emphasis i.e. density on regions for which $t_y\approx t$ as $n$ increases (that is, the bandwidth $\epsilon_n$ decreases with $n$). Hence,  we denote the two possible ABC constructions one may consider under the joint and marginal posterior models respectively, in the SMC Samplers procedure as follows:
\begin{equation}
\begin{split}
&\pi_{J,n}(\bm{\theta}, t^{1:S}|t_y)\propto \tilde{K}_{\epsilon_n}(t_y,t^{1:S})f(t^{1:S}|\bm{\theta})\pi(\bm{\theta})\\
&\pi_{M,n}(\bm{\theta}|t_y)\propto\pi(\bm{\theta})\int_{\mathcal{T}^S}\tilde{K}_{\epsilon_n}(t_y,t^{1:S})f(t^{1:S}|\bm{\theta})dt^{1:S}
\end{split}
\end{equation}

Now the aspect of this procedure that makes it adaptive is the selection of the size of the discrepancy between $\pi_n$ and $\pi_{n+1}$, for each $n \in \left\{1,2\ldots,T\right\}$ as well as the final stopping point. In this paper we propose to perform adaption of the ABC target distribution sequence at every step. The aim is to progressively select a sequence of distributions, online such that the discrepancy between the next distribution and the previous, as controlled by the tolerance $\epsilon_n$ sequence, is controlled by the ``fitness'' or efficiency of the particle approximation of the previous target distribution in the sequence. A good approximation would indicate that one may take a larger step whilst a poorer approximation indicates that smaller steps should be taken.

Formally, we perform the adaption such that a new tolerance $\epsilon_n$, at iteration $n$, is generated through a particle system based quantile matching procedure. The procedure adopted considers the new tolerance to be obtained as the solution for $\epsilon$ in the following equation
\begin{equation}
\widehat{q}_{n-1}=\epsilon \sqrt{2} \text{erf}^{-1} \left[ 2 F \left(q\right) -1 \right ], 
\end{equation}
where $q$ is a user specified quantile level, $F$ is the CDF of a normal distribution with mean 0 and standard deviation $\epsilon$ and $\widehat{q}_{n-1}$ is the particle population quantile estimate obtained from the ABC posterior approximation after correction stage in the SMC Sampler ABC algorithm. In this way the tolerance schedule is continually adapting to the local particle approximation performance. In practice it is computationally efficient to employ the following adaptive schedule for the tolerance in the ABC posterior, where we ensure that the sequence of distributions is designed such that the new tolerance is calculated as a strictly decreasing schedule given by
\begin{equation}
\epsilon_{n}=\min \left( (1-\alpha)\epsilon_{n-1},\epsilon^{\ast} \right ),
\end{equation}
where $\alpha \in \left(0,1\right)$.

\subsubsection{Choice of Mutation Kernel}
\label{sec:mutker}
There are many choices for mutation kernel that could be considered when designing an SMC Sampler algorithm. The choice of kernel is often critical to select in order for the algorithm to perform well. In this section, we first survey a few possible choices and then we present a specialized choice of mutation kernel we adopted from the genetic search literature \citep{li2009multiobjective} which involves combination of mutation and cross-over operators for the particle mutation in the SMC Sampler. In order to utilise this class of mutation operator in SMC Sampler settings we had to formally write down not just the mutation and cross-over operators in a structural form, as typically specified in the NGSAII class of genetic optimization algorithms, but to also define their distributional form. We provide these at the end of the section.

Some examples of possible choices of the mutation kernel are given as follows:
\begin{enumerate}
\item{\textbf{Independent kernels}. In this setting, one would select a mutation kernel given for all $n \in \left\{1,2,\ldots,T\right\}$ by $M_{n}\left( \bm{\theta}_{n-1},\bm{\theta}_{n}\right) = M_{n}\left( \bm{\theta}_{n}\right) $;}
\item{\textbf{Local Random Walks}. In this setting, the kernel would be selected for all\break $n \in \left\{1,2,\ldots,T\right\}$ to be of the form $M_{n}\left( \bm{\theta}_{n-1},\bm{\theta}_{n}\right) $, where the mutation from $\bm{\theta}_{n-1}$ to $\bm{\theta}_{n}$ follows a local Random Walk based around, say, a Gaussian smoothing kernel as given by \cite{givens1996local};
}
\item{\textbf{Markov chain Monte Carlo Kernels}. In this setting, the kernel would be selected for all $n \in \left\{1,2,\ldots,T\right\}$ to be an MCMC kernel of invariant distribution $\pi_n$. As noted by
\cite{del2006sequential} and \cite{peters2005topics}, this option is suitable if the Markov chain kernel is mixing rapidly or if the sequence of distributions is such that $\pi_{n-1}$ is close to $\pi_n$, which is often the case by design. Then the use of an MCMC kernel would result in running for each stage, $N$ inhomogeneous Markov chains. Then one must correct for the fact that one is not targeting the correct distribution under these Markov chains, which is achieved using IS: $\hat{\pi}_{n-1}^{N} = \sum_{i=1}^N W^{(i)}_{n-1}\delta_{\bm{\theta}_{n-1}^{(i)}}\left(\bm{\theta}\right)$ and running $L$ iterations of the Markov chain for each particle, where each of the $N$ chains will target $\sum_{i=1}^N W^{(i)}_{n-1}\prod_{l=1}^L M_{l}\left( \bm{\theta}_{l-1}^{(i)},\bm{\theta}_{l}\right)$, which is not in general $\pi_n$, then with an IS correction, such an approach is accurate and unbiased (i.e., targets the distribution of interest at time $n$ given by $\pi_n$;}
\item{\textbf{Gibbs Sampler kernels}. If the sequence of target distributions $\left\{\pi_n\right\}_{n\geq 0}$ is such that its support is multivariate, then it may also be possible to sample from the full conditional distributions in the sequence of distributions. This approach allows one to undertake a Gibbs step, which would involve a kernel for update of the $k$-th element given in the form
\begin{align}
M_{n}\left(\bm{\theta}_{n-1},d\bm{\theta}_{n}\right)  = \delta_{\bm{\theta}_{n-1,-k}}\left(d\bm{\theta}_{n,-k}\right) \pi_n\left(\bm{\theta}_{n,k}|\bm{\theta}_{n,-k}\right)
\end{align}
with $\bm{\theta}_{n,-k}= \left(\bm{\theta}_{n,1},\bm{\theta}_{n,2},\ldots,\bm{\theta}_{n,k-1},\bm{\theta}_{n,k+1},\ldots,
\bm{\theta}_{n,J}\right)$, where there are $J$ parameters in the OpRisk model target posterior. If
the full conditionals are not available, one could approximate them accurately at each stage and
then correct for the approximation error through~IS;}
\item{\textbf{Mixture kernels}. It is always possible to consider a mixture kernel choice given by
\begin{align} \label{EqnMixMutationKernel}
M_n\left(\bm{\theta}_{n-1},\bm{\theta}_{n}\right) = \sum_{m=1}^M \alpha_{n,m}\left(\bm{\theta}_{n-1}\right) M_{n,m}\left(\bm{\theta}_{n-1},\bm{\theta}_{n}\right),
\end{align}
with $\alpha_{n,m}\left(\bm{\theta}_{n-1}\right) > 0$ and $\sum_{m=1}^M \alpha_{n,m}
\left(\bm{\theta}_{n-1}\right) = 1$. One special case of this type of kernel would be
an independent kernel constructed by a kernel density estimate of $ M_{n,m}\left(\bm{\theta}_{n-1},
\bm{\theta}_{n}\right) =  M_{n}\left(\bm{\theta}_{n-1},\bm{\theta}_{n}\right)$ for all $m$
and $\alpha_{n,m}\left(\bm{\theta}_{n-1}\right) = W_{n-1}^{(i)}$ with $M=N$; }
\item{\textbf{Partial Rejection Control kernels}. In this case, one aims to construct a mutation kernel in the SMC Sampler that guarantees all sampled particles have importance weights with a ``fitness'' exceeding a user-specified threshold at each time $n$, denoted by $c_n$ such that $w_n^{(i)} \geq c_n, \; \forall i \in \left\{1,2, \ldots, N\right\}$. To achieve this, one modifies any of the earlier mutation kernels to take the form given by
\begin{align}
\label{eqn:emstar}
    M_{n}^{\ast }\left( \bm{\theta}^{i}_{n-1},\bm{\theta}_{n}\right) =r(c_{n}, \bm{\theta}_{n-1}^{(i)})^{-1} \left[ \min\left\{ 1, W_{n-1}^{(i)}\frac{w_n\left(
    \bm{\theta}_{n-1}^{(i)},\bm{\theta}_{n}\right) }{c_{n}}\right\} M_{n}\left(
    \bm{\theta}_{n-1}^{(i)},\bm{\theta}_{n}\right)\right].
\end{align}
The quantity $r(c_{n}, \bm{\theta}_{n-1}^{(i)})$ denotes the normalizing constant for particle $\bm{\theta}_{n-1}^{(i)}$, given by
\begin{align}\label{Rejection Probability}
r(c_{n}, \bm{\theta}_{n-1}^{(i)})=\int \min \left\{ 1, W_{n-1}^{(i)}\frac{w_n\left( \bm{\theta}_{n-1}^{(i)},\bm{\theta}_{n}\right) }{c_{n}}\right\} M_{n}\left( \bm{\theta}_{n-1}^{(i)},\bm{\theta}_{n}\right) d\bm{\theta}_{n}.
\end{align}
Note that $0<r(c_n,\bm{\theta}_{n-1})\leq 1$ if  (w.l.o.g.) the mutation kernel $M_n$ is normalized, so that $\int M_{n}(\bm{\theta}_{n-1},\bm{\theta}_n)d\bm{\theta}_n=1$, and if the PRC threshold $0\leq c_n<\infty$ is finite. The sequence of PRC thresholds is then user-specified to ensure a certain particle ``fitness'' at each stage of the SMC Sampler. We will detail more explicitly this example in a future section.
}
\item{\textbf{Genetic Mutation and Cross-Over operators}. In this class of mutation kernel in the SMC Sampler we consider the class of genetic algorithm type mutations. In particular, we describe the class of MOEA mutation and crossover operators widely used in stochastic search algorithms, introduced in the method of \cite{deb2002fast}. This class of mutation kernel is the most widely used operator in multi-objective optimisation and we demonstrate its adaption to the SMC Sampler framework. 

A disadvantage of the NSGA-II operators is that they are only able to mutate binary, integer, or real encodings of the output parameter vectors, whereas the stochastic process for the limit order submission activity by liquidity providers requires the specification of a positive definite and symmetric covariance matrix for the generation of intensities from a multivariate skew-t distribution. The positive definiteness and symmetry constraints of the covariance matrix will not be preserved if one simply employs the evolutionary operators above to produce new sets of covariance matrix candidate solutions. For this reason, \cite{panayi2015stochastic} introduced a new covariance mutation operator which generates new candidate covariance matrices which remains in the manifold of positive definite matrices.

\textbf{Simulated Binary Crossover (SBX):} From two previous particles $\bm{\theta}_{n-1}^{(i)},\bm{\theta}_{n-1}^{(j)}$, a new solution $\bm{\theta}_{n}^{(i)}$ is formed, where the $k$-th elements is crossed as follows:
\begin{align}
\theta_{n}^{(i,k)} &=\frac{1}{2}[(1-\bar{\beta})\theta_{n-1}^{(i,k)}+(1+\bar{\beta})\theta_{n-1}^{(j,k)}].
\end{align}
Here, $\bar{\beta}$ is a random sample from a distribution with density 
\begin{equation*}
\bar{\beta} = \begin{cases} (\alpha u)^\frac{1}{\eta_c +1},  &\mbox{if }u \leq \frac{1}{\alpha}  \\ 
(\frac{1}{2-\alpha u})^\frac{1}{\eta_c +1}, & \mbox{otherwise}. \end{cases} 
\end{equation*}
where $u \sim U(0,1)$ and $\alpha=2-\beta^{-(\eta_c+1)}$, with 
\begin{equation}
\beta=1+\frac{2}{\theta_{n-1}^{(j,k)}-\theta_{n-1}^{(i,k)}}\min \left[\left( \theta_{n-1}^{(i,k)}-\theta_{k_{L}}  \right),\left(\theta_{k_{U}}-\theta_{n-1}^{(j,k)}\right) \right]
\end{equation}

This would produce a mutation kernel for this type of move at SMC Sampler iteration $n$ for the $k$-th element of the $i$-th particle vector which would be updated according to a density given by
\begin{align*}
M_n(\theta_{n}^{(i,k)}|\theta_{n-1}^{(i,k)},\theta_{n-1}^{(j,k)})
&=\frac{2}{\theta_{n-1}^{(j,k)}-\theta_{n-1}^{(i,k)}} 
\left[ \mathbb{1}_{\theta_{n}^{(i,k)} \in \left[ \theta_{n-1}^{(i,k)}, \frac{\theta_{n-1}^{(i,k)}+\theta_{n-1}^{(j,k)}}{2} \right ] }\left( \frac{\eta_c+1}{\alpha}\cdot\frac{\theta_{n-1}^{(i,k)}+\theta_{n-1}^{(j,k)} - \theta_{n}^{(i,k)}}{\theta_{n-1}^{(j,k)}-\theta_{n-1}^{(i,k)}} \right )^{\eta_c}  \right.\\
& + \left. \mathbb{1}_{\theta_{n}^{(i,k)} \in \left[ \frac{\theta_{n-1}^{(i,k)}+\theta_{n-1}^{(j,k)}}{2} - \frac{1}{2(2-\alpha)^{\frac{1}{\eta_c +1}}},\theta_{n-1}^{(i,k)}\right]} 
\left( \frac{\eta_c+1}{\alpha}\cdot\frac{\theta_{n-1}^{(i,k)}+\theta_{n-1}^{(j,k)} -\theta_{n}^{(i,k)}}{\theta_{n-1}^{(j,k)}-\theta_{n-1}^{(i,k)}} \right )^{\eta_c+2}  
\right ]
\end{align*}

We use the crossover operator with probability $p_c=0.7$ and a distribution index $\eta_c=5$. Every element $k$ of the $i$-th particle vector is crossed with probability 0.5.  

\textbf{Polynomial mutation:} The mutation operator perturbs elements of the solution, according to the distance from the boundaries. 
\begin{equation*}
\theta_{n}^{(i,k)}=\theta_{n-1}^{(i,k)}+\bar{\delta}(\theta_{k_{U}}-\theta_{k_{L}})
\end{equation*}
where we have for $\bar{\delta}$
\begin{equation*}
\bar{\delta}=\begin{cases} \left[2\gamma+(1-2\gamma)(1-\delta)^{\eta_m+1}\right]^{\frac{1}{\eta_m+1}}-1 &\mbox{if }\gamma<0.5  \\ 
1-\left[2(1-\gamma)+2(\gamma-0.5)(1-\delta)^{\eta_m+1}\right]^{\frac{1}{\eta_m+1}} & \mbox{if } \gamma \geq 0.5. \end{cases} 
\end{equation*}
with 
\begin{equation*}
\delta=\min \left[\left( \theta_{n}^{(i,k)}-\theta_{k_{L}}  \right),\left(\theta_{k_{U}}-\theta_{n}^{(i,k)}\right) \right].
\end{equation*}
where, $\gamma \sim U(0,1)$. 

This would produce a mutation kernel for this type of move at SMC Sampler iteration $n$ for the $k$-th element of the $i$-th particle vector which would be updated according to a density given by
\begin{align*}
M_n(\theta_{n}^{(i,k)}|\theta_{n-1}^{(i,k)})=\frac{1}{\theta_{k_{U}}-\theta_{k_{L}}} \left[
\mathbb{1}_{\theta_{n}^{(i,k)} \leq \theta_{n-1}^{(i,k)}}\left( \frac{ (\eta_m+1)(\bar{\delta}+1)^{\eta_m} }{2(1-(1-\delta)^{\eta_m+1})}  \right ) + \right.\\
\left. \mathbb{1}_{\theta_{n}^{(i,k)} > \theta_{n-1}^{(i,k)}}\left( \frac{ (\eta_m+1)(1-\bar{\delta})^{\eta_m} }{2(1-(1-\delta)^{\eta_m+1})}  \right )
 \right ]
\end{align*}

The distribution index $\eta_m=10$. The polynomial mutation operator is used with probability $p_m=0.2$.

\textbf{Covariance mutation operator:} In the $t$-th generation of the MOEA, we generate $\left \{ \Sigma_{t}^{(i)} \right \},i=1\ldots N$ from a mixture distribution $M_n(\Sigma_{n,i})$ defined as follows:
\begin{equation*}
M_n(\Sigma_{t}^{(i)})=(1-w_1)\mathcal{IW}(\Psi_n,p_1)+w_1 \mathcal{IW}(\Psi,p_2)
\end{equation*} 
where $\mathcal{IW}$ denotes the Inverse Wishart distribution, $p_1,p_2$ are degrees of freedom parameters with $p_2<p_1$, and where $w_1$ is small so that sampling from the second distribution happens infrequently. Here $\Psi$ denotes an uninformative positive definite matrix, with the effect that sampling from the second distribution leads to moves away from the local region being explored. $\Psi_t$ is also a positive definite matrix, fitted based on moment matching to the sample mean of the successfully proposed candidate solutions in the previous stage of the Multi-Objective optimisation as follows:
\begin{equation*}
\Psi_n=\frac{1}{\sum_{s=1}^n w^{s}} 
\sum_{s=1}^n w^{s} \frac{1}{\sum_{i=1}^N \frac{1}{r_{s}^{(i)}}} 
\sum_{i=1}^N \frac{1}{r_{s}^{(i)}}  \tilde{\Sigma}_{n}^{(i)}
\end{equation*} 
where $r_{s}^{(i)}$ is the non-domination rank of the $i$-th solution in the $s$-th generation, and $w^s$ with $w<1$ is an exponential weighting factor. 
}
\end{enumerate}

\section{Application to equities LOB data: data description}

The data employed in this study constitutes the intra-day trading activity on the European multi-lateral trading facility (MTF) Chi-X Europe between January and April 2012. Chi-X Europe operated as an individual entity from 2007, before being purchased by BATS Europe at the end of the trading period under consideration. We note that Chi-X Europe is a secondary exchange, i.e. the securities that are traded on the exchange are listed and primarily traded on national/supranational exchanges, including the London Stock Exchange, Euronext, Deutsche Boerse, and the SIX Swiss Exchange, amongst others. However, it maintains a significant proportion of the daily trading activity in each of these markets, between 20\% and 35\% in most cases\footnote{\url{http://www.liquidmetrix.com/liquidmetrix/battlemap}}.

\begin{figure}[ht]
 \begin{center}
 \includegraphics[width=0.55\textwidth]{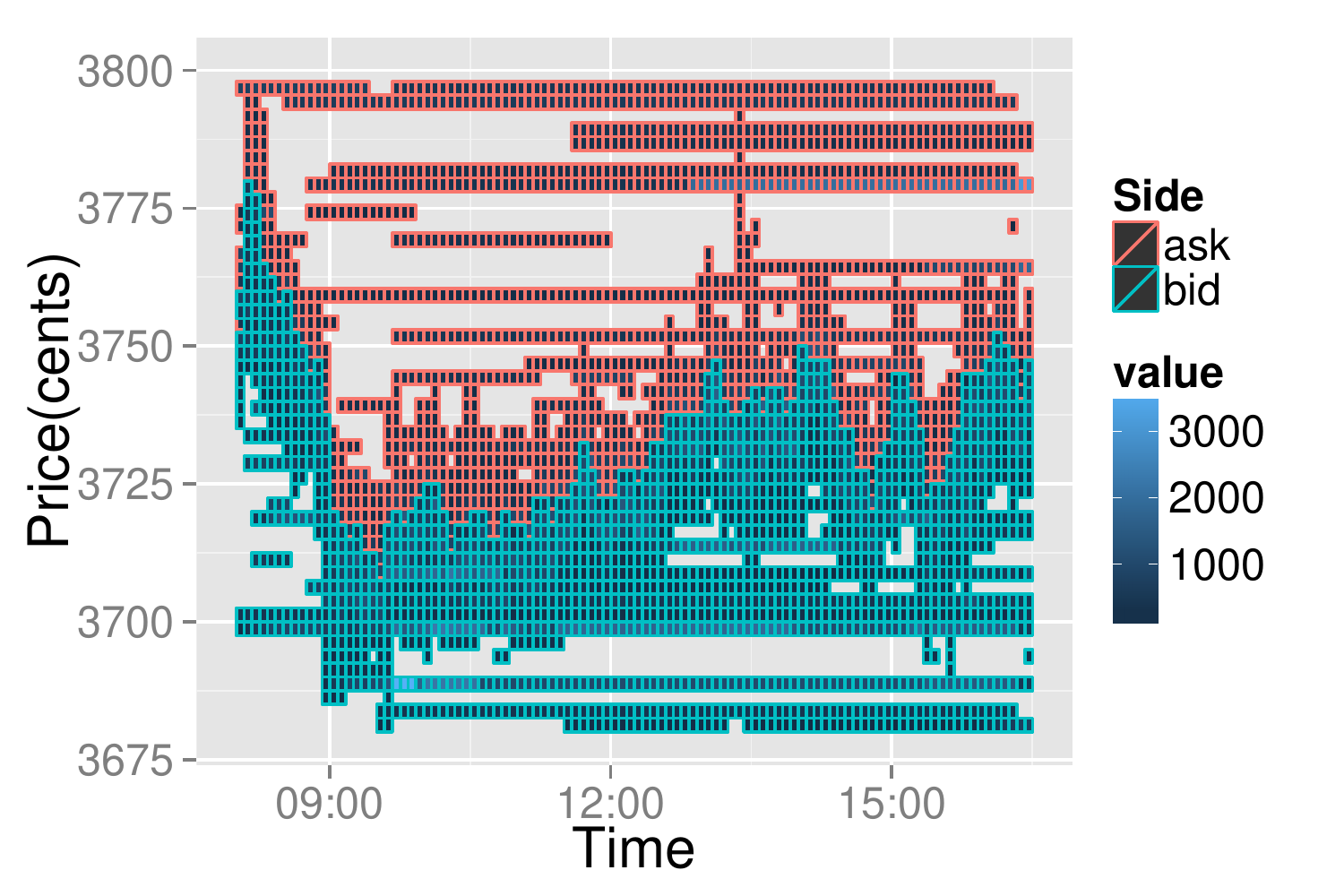}
  \includegraphics[width=0.44\textwidth]{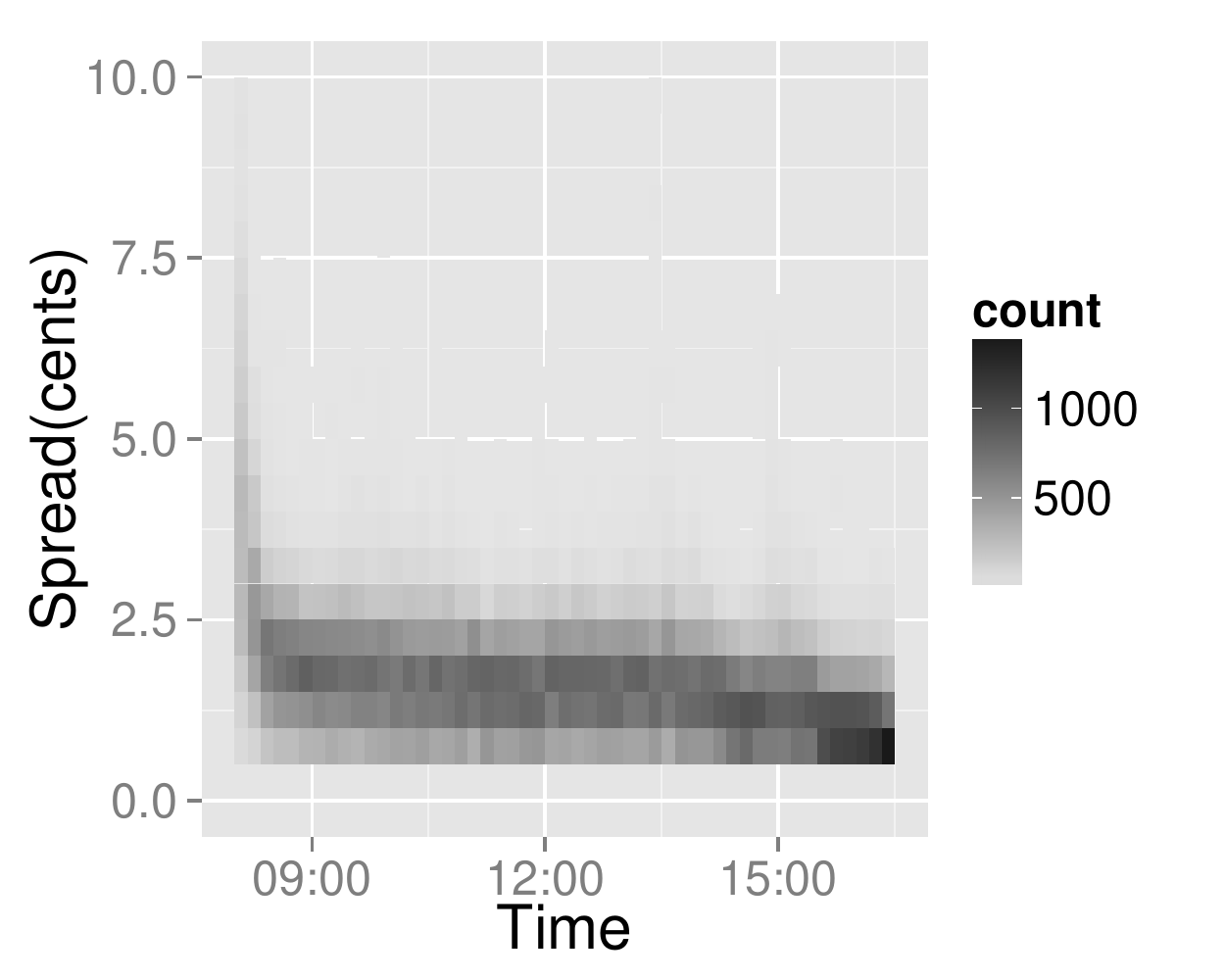}
 \caption{(Left): A representation of real market data intra-day LOB states obtained from the trading activity for asset BNP Paribas SA on the 5th of March 2012. The shading of each box indicates the volume available at that price, which is volume available to buy for blue-bordered boxes and volume available to sell for red-bordered boxes. (Right): A heatmap of the intra-day spread for the period February to March 2012 for asset BNP Paribas SA.}
 \label{fig:LOBandspread}
 \end{center}
 \end{figure}

The complete dataset covers over 1300 assets, primarily stocks, but also including exchange-traded funds (ETFs) and American depositary receipts (ADRs). For the purposes of this study, we select one of the most commonly traded stocks in the French CAC 40 Index, namely BNP Paribas SA. Figure \ref{fig:LOBandspread} shows the evolution of the LOB on a typical day for this asset based on real market observation data from the LOB. We also present a heatmap of the inside spread $S_t=P_t^{a,1}-P_t^{b,1}$ over the 2 month period February to March 2012. The inside spread is the most common measure of `liquidity', i.e. the relative ease with which one can buy or sell a financial asset.

Chi-X Europe operates both a visible and a hidden order book, and traders have the option to route orders to the hidden book if they meet certain conditions relating to order type and size. The dataset consists of only data in the visible book, after it has been processed by the exchange's matching engine. That is, while the exchange allows for a range of order types with time-in-force modifiers, the processed data consists of the timestamps and order sizes of limit order submissions, executions and cancellations. However, this data is sufficient to construct a much more detailed picture of the state of the LOB than is typically available in previous studies (which only consider aggregate volume in either the first level or the first 5 levels), as we can disaggregate the volumes per level up to any depth in the LOB.      

The raw, unevenly spaced data is thus used to construct the state of the LOB at each event timestamp (these are accurate up to millisecond precision). Because of our interest in fitting the auxiliary models describing price and volume dynamics (these are outlined in Section \ref{sec:results}), however, we subsample the process at regular 10 second intervals, in order to then extract the price and volume variables of interest. Thus, from an irregularly spaced process typically containing between 50,000 and 500,000 events every day, we extract a regular timeseries of the auxiliary model variables for the purposes of our estimation. 

\section{Results}
\label{sec:results}

The results presented in this section may be compared to those obtained from indirect inference procedures as reported in \cite{panayi2015stochastic}. To achieve comparison we have also provide the results for what they termed the benchmark `reference' model, which makes a series of assumptions in order to simplify estimation and model structure. This basic reference model has the following parameter vector $\left \{ \mu_0^{LO,p},\mu_0^{LO,d}, \mu_0^{MO},\gamma_0,\nu, \sigma^{MO} \right \}$, as well as the covariance matrix $\Sigma$ to be estimated, see details in \cite{panayi2015stochastic}. The results will be presented in terms of the ABC marginal posterior distributions of the individual parameters of the LOB simulation and in terms of the resulting stochastic agent based LOB model to reasonably produce realistic features of the simulated LOB intra-daily. 

In our introduction to likelihood-free methods in Section \ref{section:abc}, we discussed the reduction of the observed data $\bm{y}$ to a low-dimensional vector of summary statistics $t_{y}$. We are interested specifically in two of the most commonly studied LOB characteristics, which correspond to the volatility in the log returns obtained from the price process dynamic (as obtained from half the inside spread) and the evolution of the volume resting on the LOB (as measured by the instantaneous aggregate total volume on the bid and ask at levels 1 to 5). The summaries we adopt at this stage are less-standard in ABC applications since they employ a functional (i.e. regression model based) summary of features of observable LOB process. In this case the summary information becomes the model characterization (dimensional reduction) captured by the estimated model parameters fit to the real LOB data and the simulated LOB data for price or volume dynamic. Specifically we have:

\noindent \textbf{Auxiliary Model 1 - Price features:} If we denote the mid-price as $p^{mid}_t=\frac{p^{a,1}_t+p^{b,1}_t}{2}$ then the log return is defined as 
\begin{equation*}
r_t=\ln \frac{p^{mid}_t}{p^{mid}_{t-\Delta_t}}
\end{equation*}
where $\Delta_t$ is a suitable interval, in our case 1 minute. We fit a GARCH(1,1) model for this aspect of the data parameterized by $\widehat{\bm{\beta}}_{1}$.

\noindent \textbf{Auxiliary Model 2 - Volume features:} We fit an MA(1) model to the detrended total volume (i.e. an ARIMA(0,1,1) model) in the first 5 levels on both the bid and ask side parameterized by $\widehat{\bm{\beta}}_{2}$, in order to capture the time series structure of the LOB volumes. 

The auxiliary models are fit to both the real and simulated data, and for the distance we estimate the Euclidean distances between the auxiliary parameter vectors
\begin{align*}
\mathcal{D}_1&=\mathcal{D}\left(\widehat{\bm{\beta}}_{1}\left( \bm{y}\right),\widehat{\bm{\beta}}_{1}( \bm{y}^\ast(\bm{\theta})) \right),\\
\mathcal{D}_2&=\mathcal{D}\left(\widehat{\bm{\beta}}_{2}\left( \bm{y}\right),\widehat{\bm{\beta}}_{2}( \bm{y}^\ast(\bm{\theta})) \right).
\end{align*}

\subsection{Estimation algorithm configuration}
\label{sec:estconfig}
To perform the estimation there are also a number of inputs to the SMC Sampler ABC algorithm that we specify, including the number of particles, the tolerance schedule forced decrement amount and the total number of iterations over which to run the estimation. Specifically, we have for our estimation procedure:
\begin{itemize}
\item The estimation procedure was run for 20 iterations.;
\item The tolerance schedule employed was the forced decrement schedules specified in Section \ref{sec:tolschedule}, with a decrement parameter $\alpha=0.1$.;
\item We obtain results using 50, 100 and 200 particles per iteration.;
\item We also tested the quality of the results for a series of quantile levels for the tolerance, i.e. $q_{0.5},q_{0.75}$ and $q_{0.9}$.
\end{itemize}  

\begin{figure}[ht]
 \begin{center}
 \includegraphics[width=0.59\textwidth]{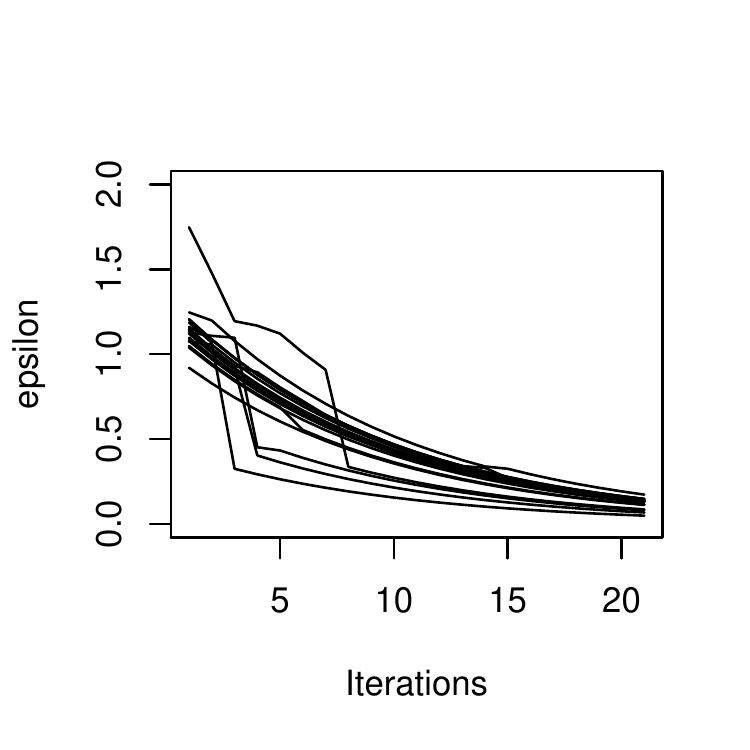}
 \caption{The adaptively estimated tolerance schedule obtained from multiple SMC Sampler-ABC runs on real data for BNP Paribas on 05/03/2012 specified in Section \ref{sec:tolschedule}. }
 \label{fig:tolschedule}
 \end{center}
 \end{figure}

Carrying out the estimation procedure for each configuration above indicated that the best results (in terms of the lowest values of $D_1, D_2$) were obtained for a quantile level $q_{0.9}$ for the tolerance, and 200 particles. We repeated the estimation procedure 20 times with this configuration and configuration above and Figure \ref{fig:tolschedule} shows the evolution of the tolerance in the ABC posterior in the case of the forced tolerance schedule, when the estimation is run for $T=20$ iterations.   

We note that the mutation operator for the covariance matrix, specified in Section \ref{sec:mutker}, which was composed of both an exploration and a mutation component, could lead to particle degeneracy in higher dimensions. Consequently, in practice it can be computationally more efficient to simplify the mutation kernel for the covariance matrix to a static mutation kernel which would eliminate the prior weighting in the numerator and denominator of each incremental particle weight. When this was performed it produces particle systems which less degeneracy issues in higher dimensions.

Secondly, due to the nature of the crossover operator, it is possible for a particle to cross with an identical particle, for example if the two particles were produced in the resampling step of the previous iteration. In our estimation we explicitly exclude this possibility and where a particle is chosen to cross with an identical particle, it is instead mutated using the operator specified in Section \ref{sec:mutker}. 

 \begin{figure}[ht]
 \begin{center}
    \includegraphics[width=0.49\textwidth]{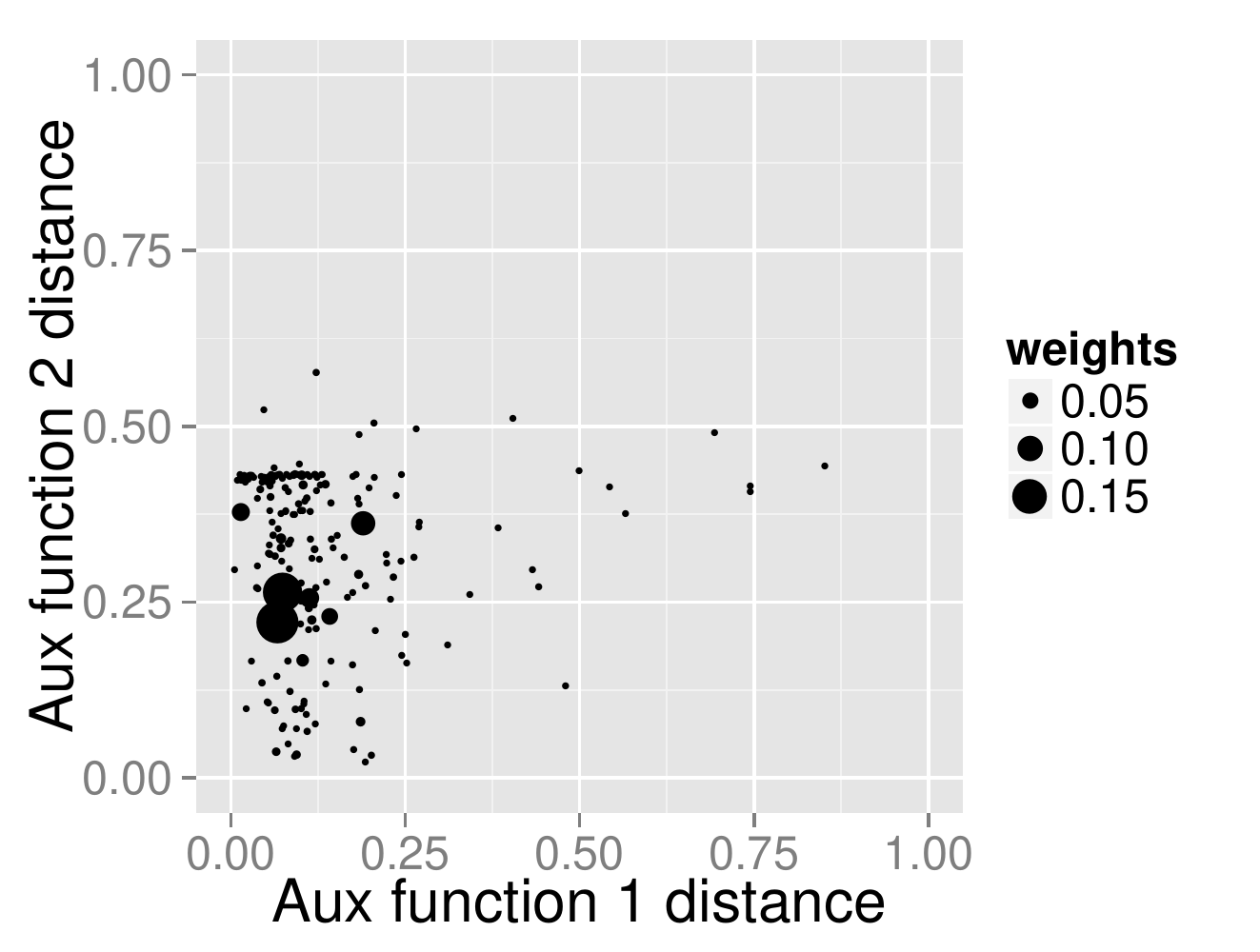}
 \includegraphics[width=0.49\textwidth]{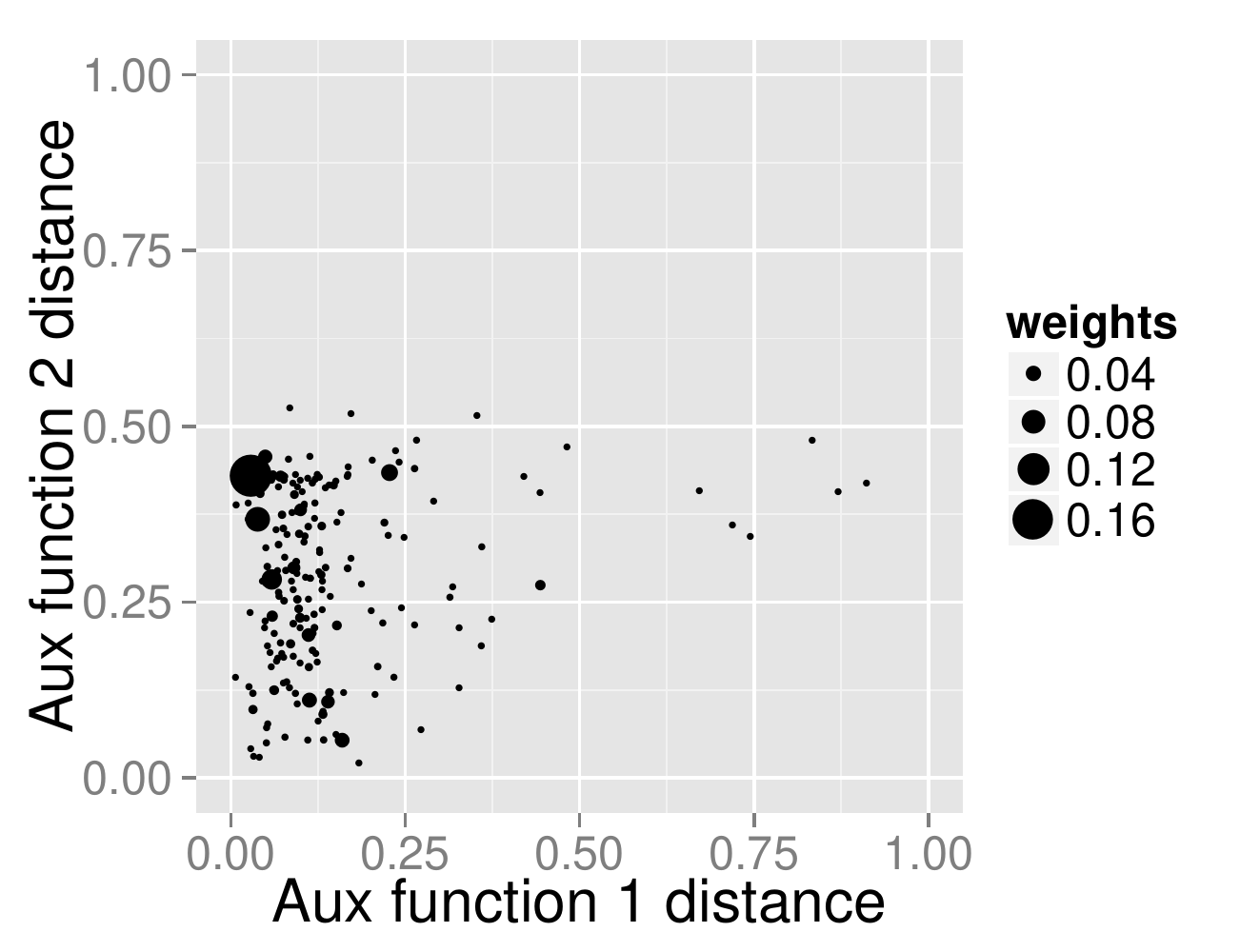}
 \caption{The realized objective function (distance metrics $\mathcal{D}_1$ and $\mathcal{D}_2$) values from each partcle in the SMC Sampler-ABC algoirthm at the final iteration for independent trials on the real data for BNP Paribas on 05/03/2012. The x-axis is the GARCH(1,1) model parameter distance discrepancies for the intra-day volatility dynamic of the price process. The y-axis is the ARIMA(0,1,1) model parameter distance discrepancies for the intra-day volume process dynamics.}
 \label{fig:multres}
 \end{center}
 \end{figure}
 
\subsection{Final particle fitness and distributions of parameters} 

Having run the SMC Sampler-ABC algorithm on the BNP Paribas LOB data for 05/03/2012 we obtained estimates of the posterior for the agent based LOB simulation model. The first set of results shows the accuracy of the LOB model to replicate features of the real LOB stochastic process relating to price and volume dynamics. This is clearly illustrated in Figure \ref{fig:multres} in terms of the values of the objective functions $\mathcal{D}_1, \mathcal{D}_2$ for each of the particles at the final iteration stage of the SMC Sampler-ABC algorithm. This is the standard way in which results for optimisation using multi-objective evolutionary algorithms (MOEAs) are presented (see discussion in \cite{panayi2015stochastic}), in order to show the Pareto optimal front that is obtained in that setting, see the discussion in the Section \ref{sec:moeaproc}. 

\begin{figure}[ht]
 \begin{center}
    \includegraphics[width=0.49\textwidth]{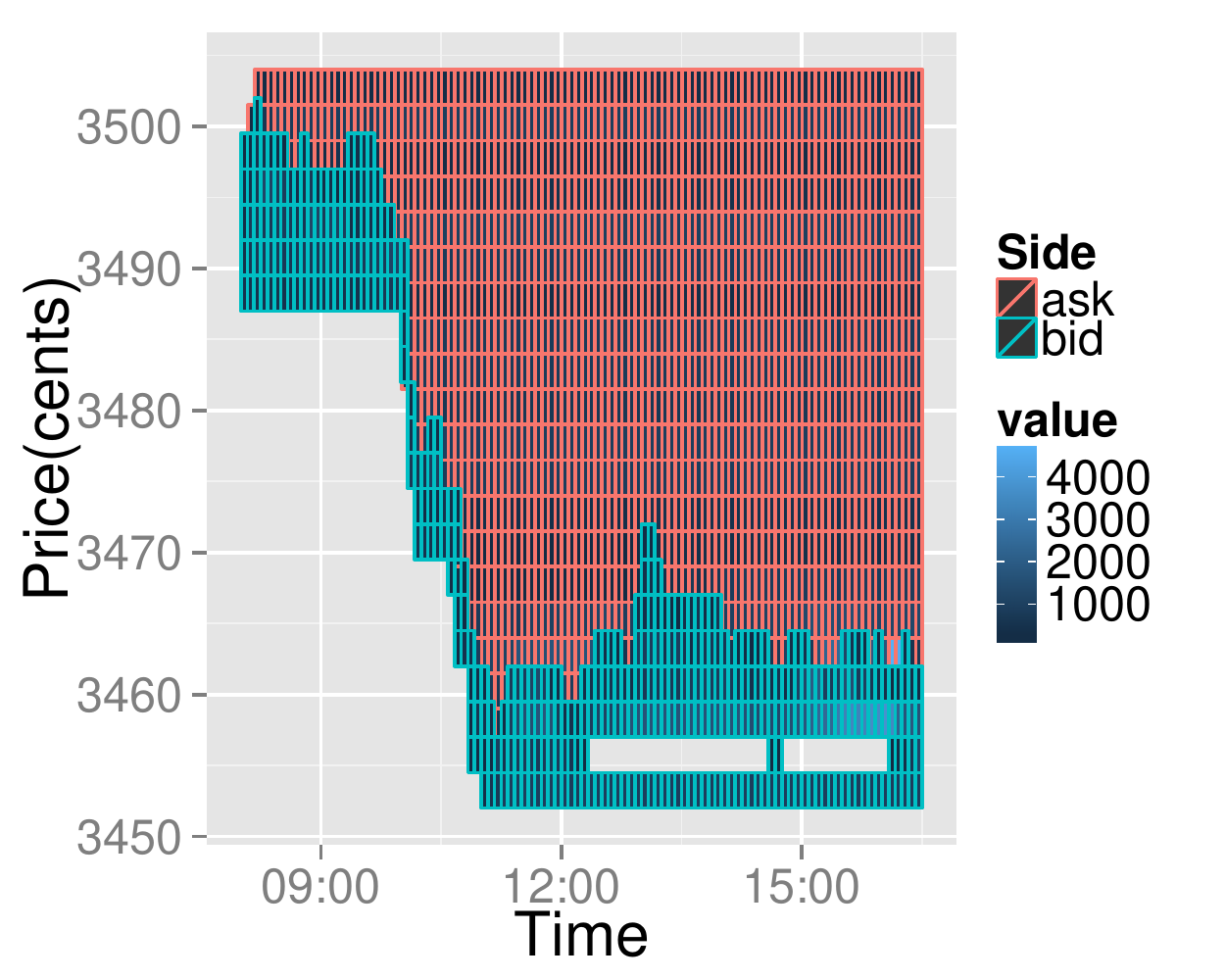}
    \includegraphics[width=0.49\textwidth]{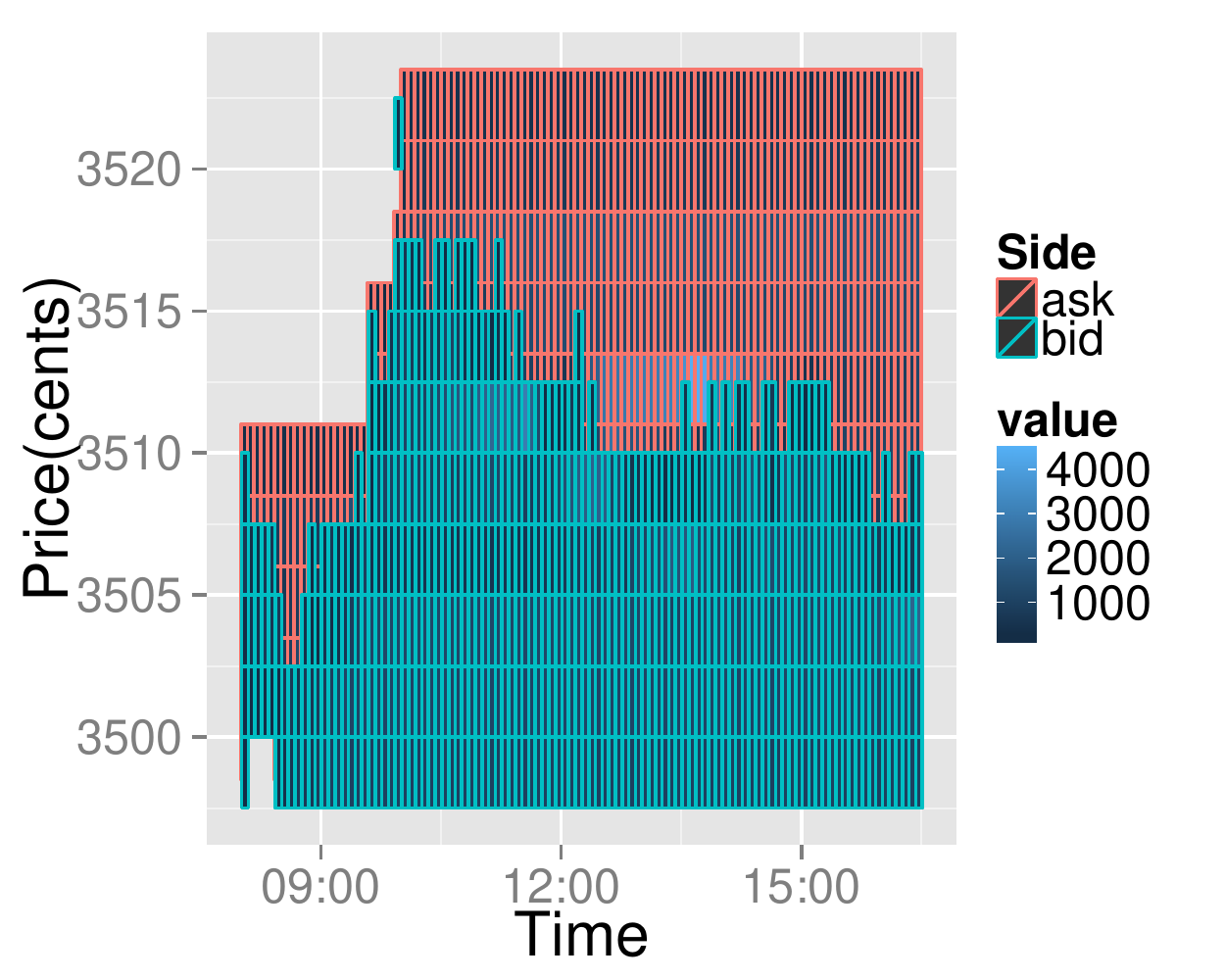}
 \includegraphics[width=0.49\textwidth]{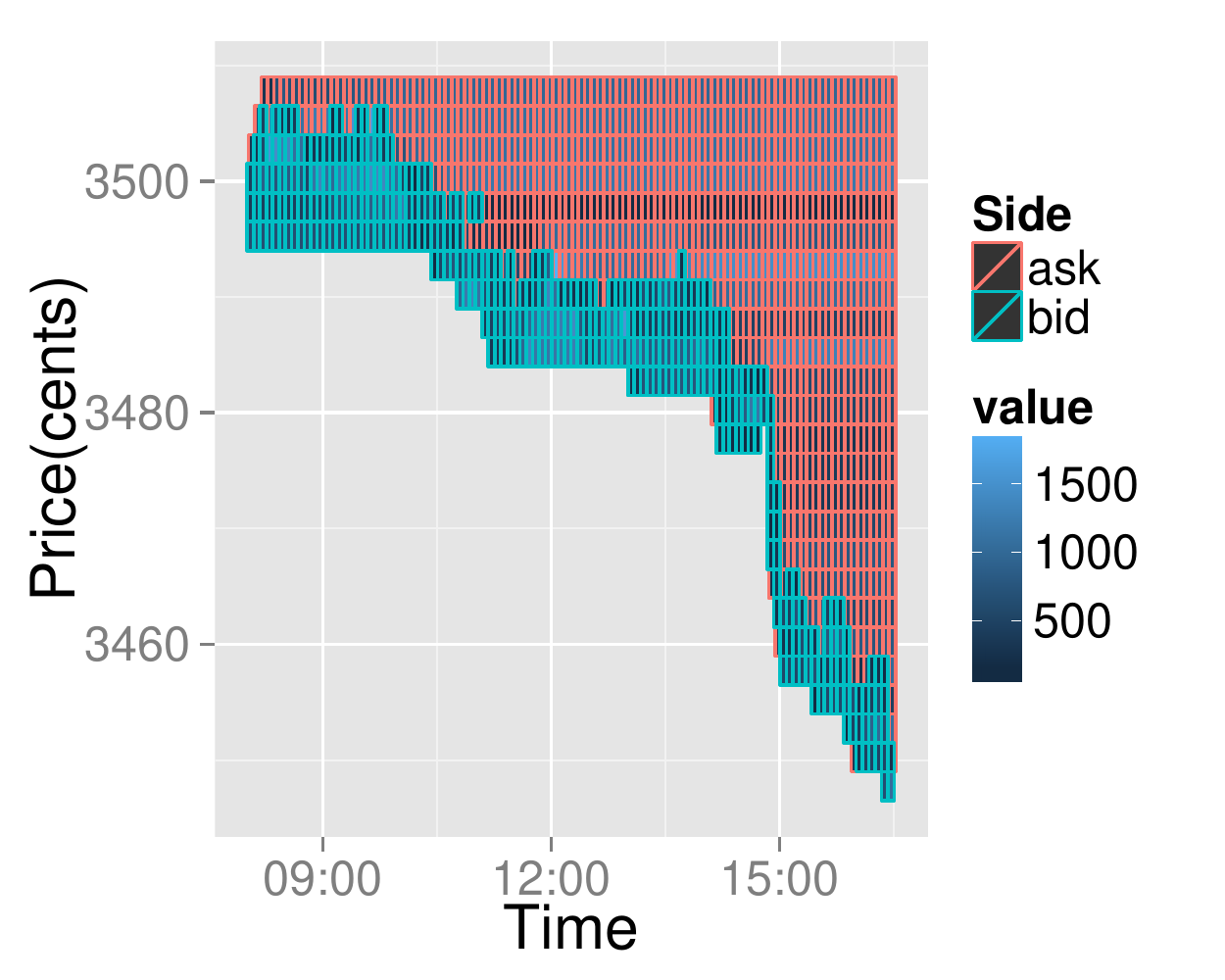}
  \includegraphics[width=0.49\textwidth]{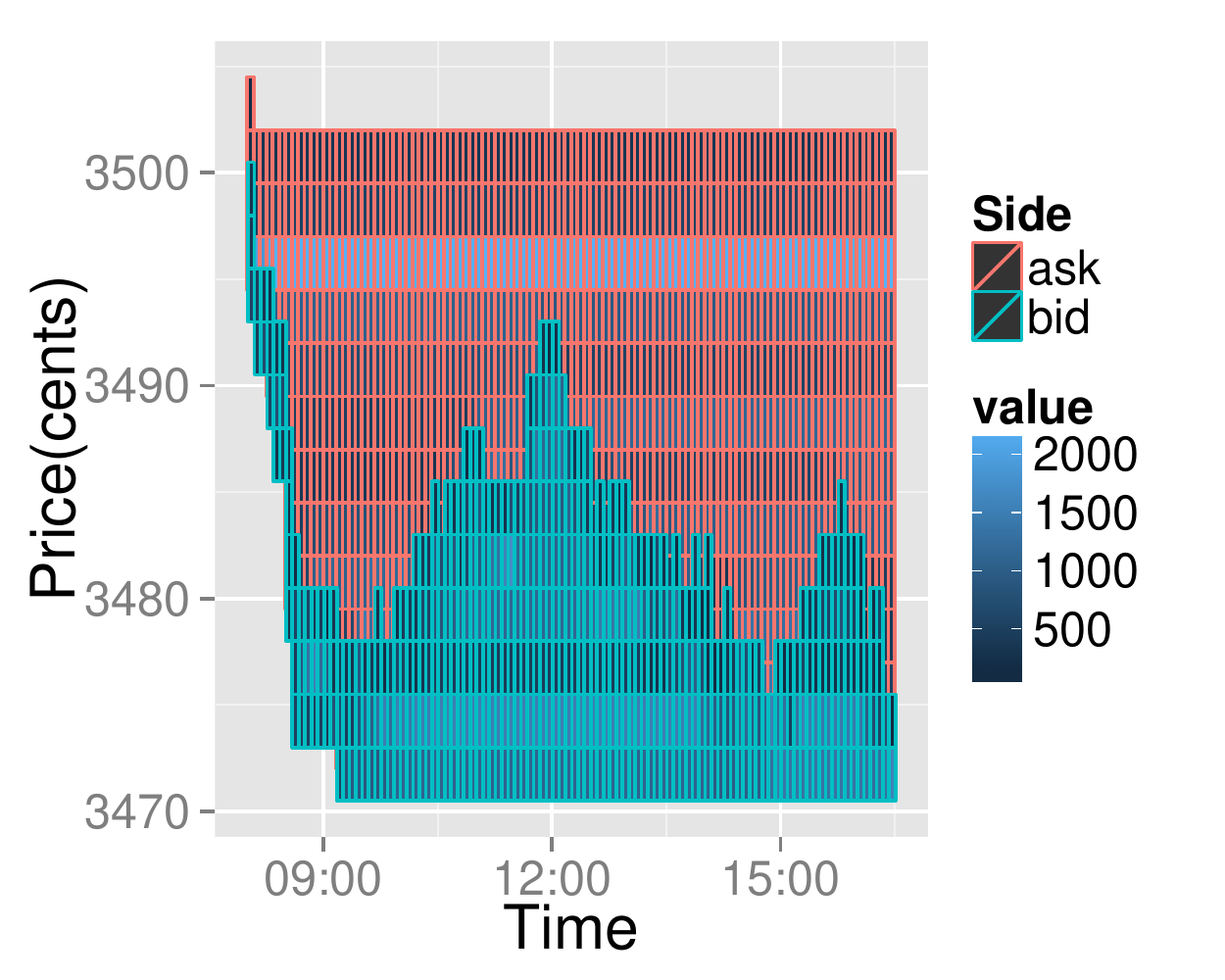}
 \caption{Representations of simulated intra-day LOB states obtained from using the (Top): MAP particle from a single estimation procedure and (Bottom): MMSE particle estimates.
 }
 \label{fig:multressim}
 \end{center}
 \end{figure}

We also present realisations of the LOB intra-day evolution for both the particle with the highest weight and the weighted mean of the particles in Figure \ref{fig:multressim}. We note that there are differences in the intra-day dynamics of the simulated financial market resulting from different repetitions of the estimation procedure. However, we note that for a subset of particles we can recover price and volume dynamics that are similar to those observed in the real market (an example of which we had seen in Figure \ref{fig:LOBandspread}).      

     \begin{figure}[ht]
 \begin{center}
    \includegraphics[width=0.49\textwidth]{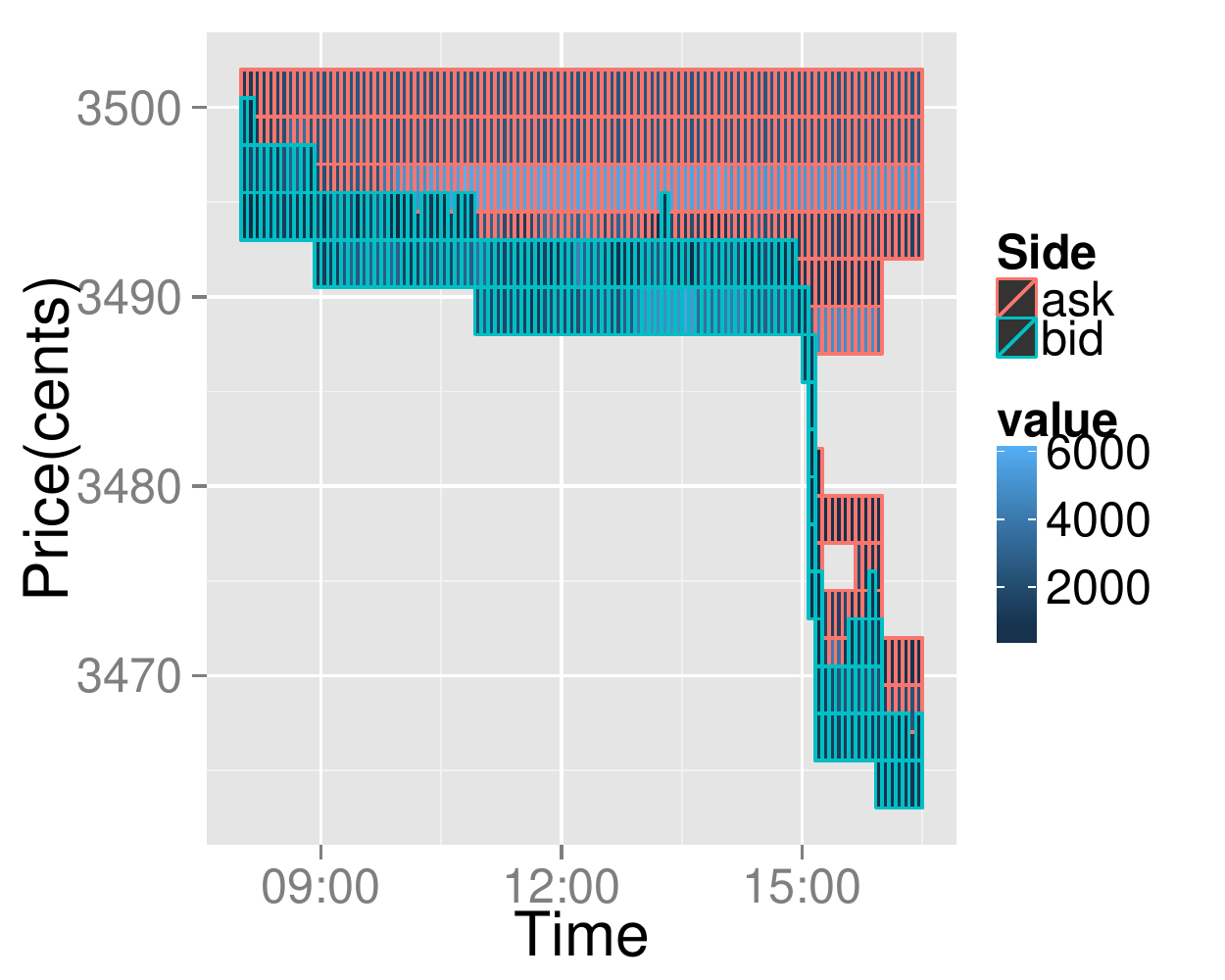}
    \includegraphics[width=0.49\textwidth]{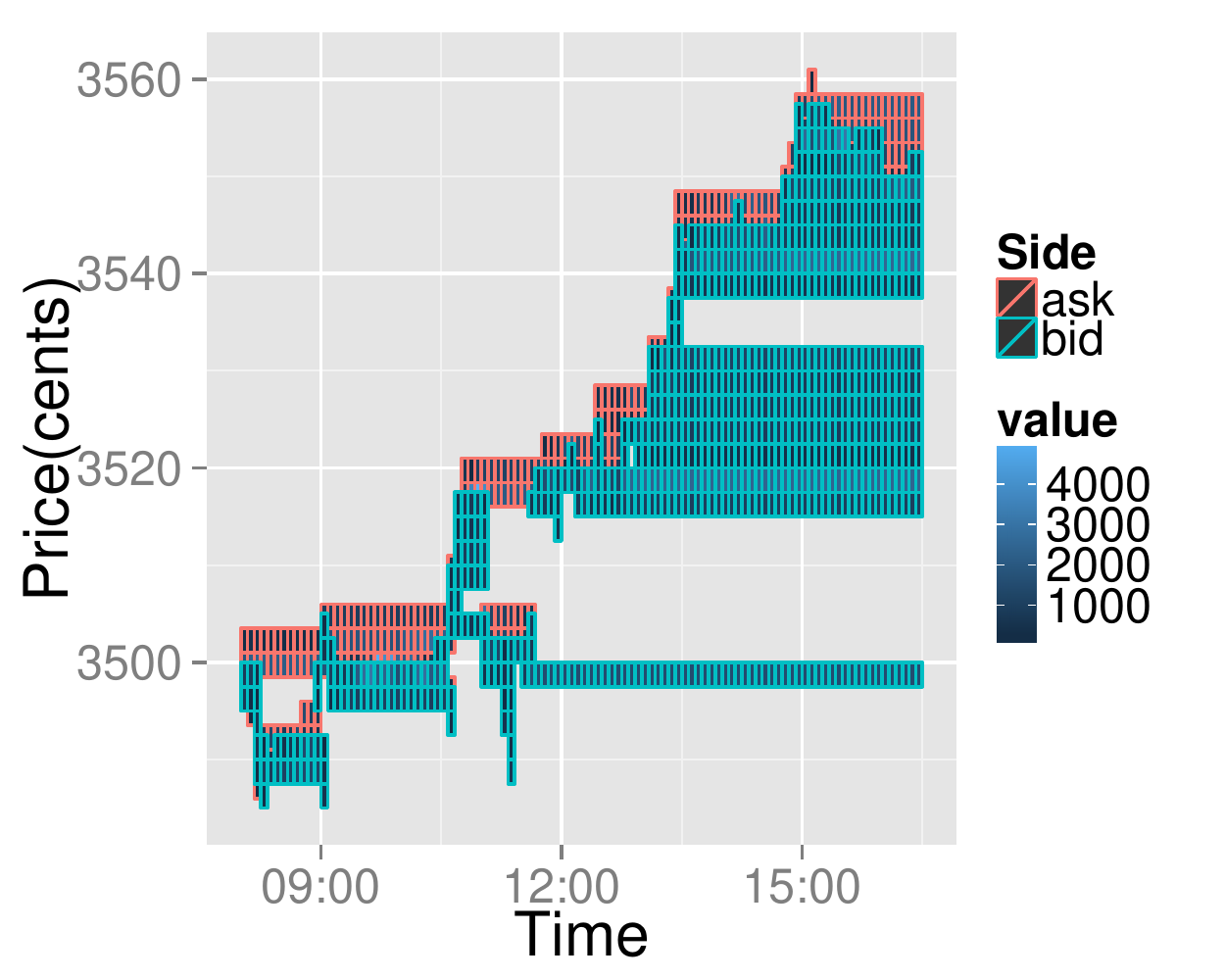}
 \includegraphics[width=0.49\textwidth]{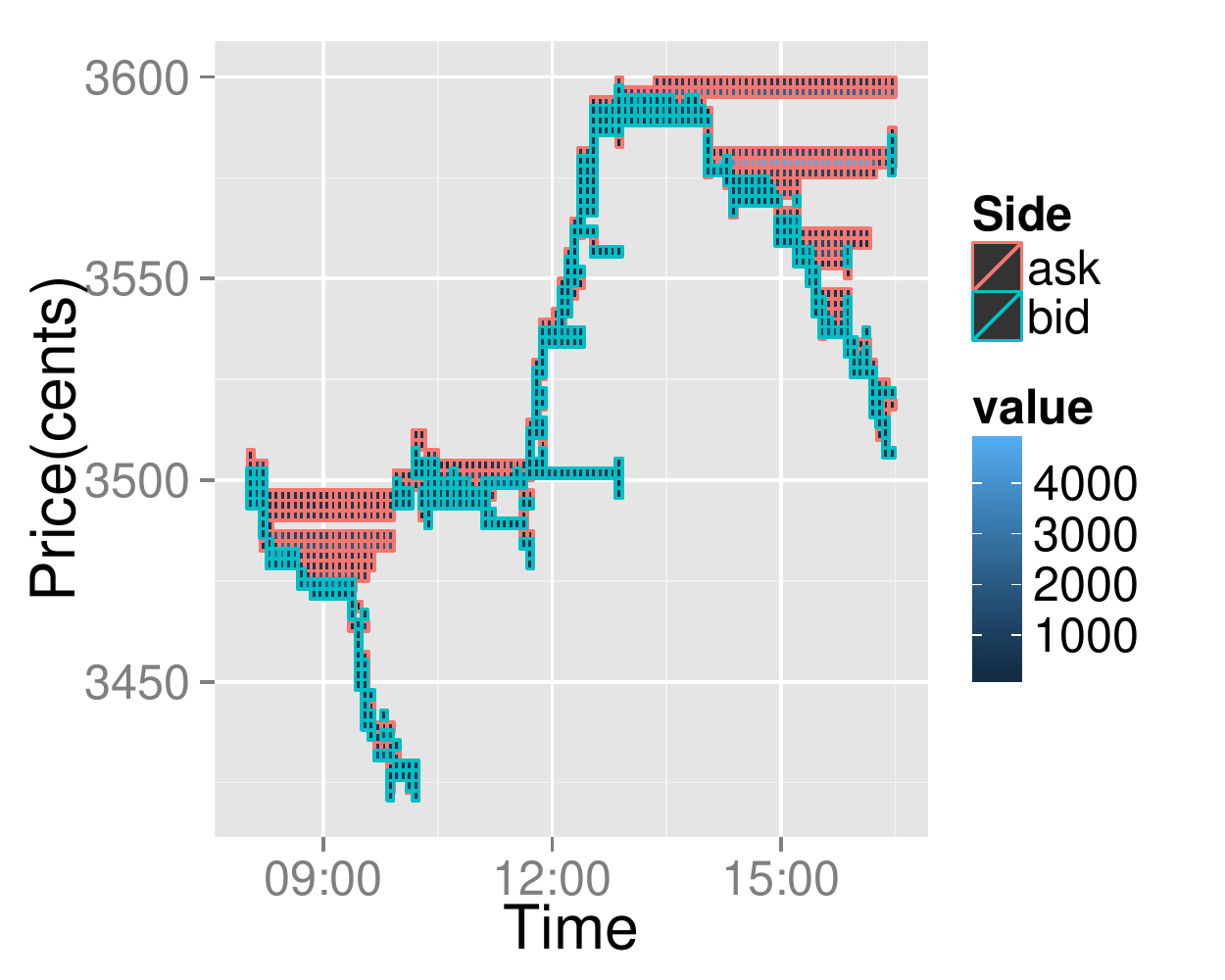}
  \includegraphics[width=0.49\textwidth]{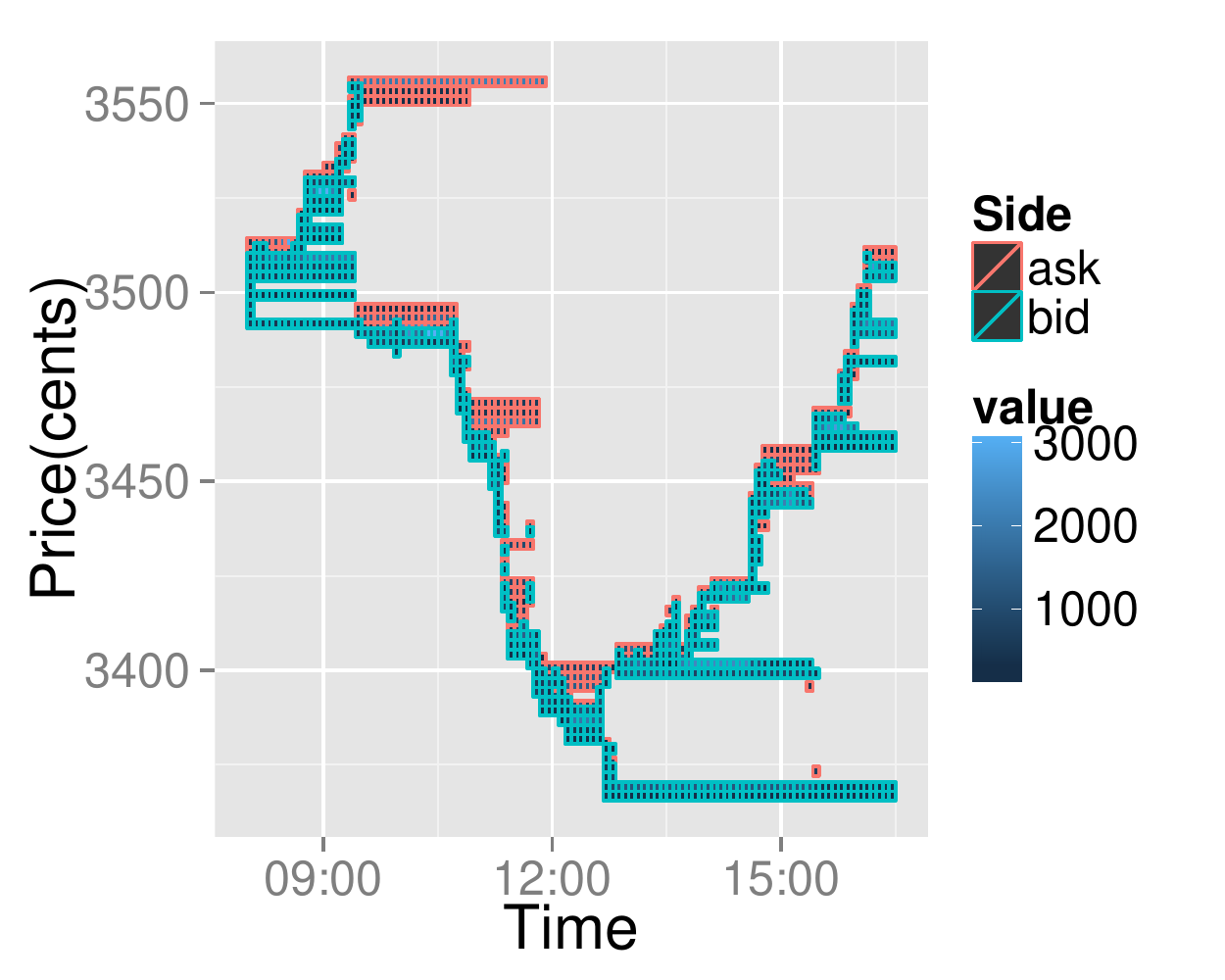}
 \caption{Representations of simulated intra-day LOB states obtained from using the (Top): MAP particle from a single estimation procedure and (Bottom): MMSE particle estimates.
 }
 \label{fig:multressim2}
 \end{center}
 \end{figure}
 
     \begin{figure}[ht]
 \begin{center}
    \includegraphics[width=0.49\textwidth]{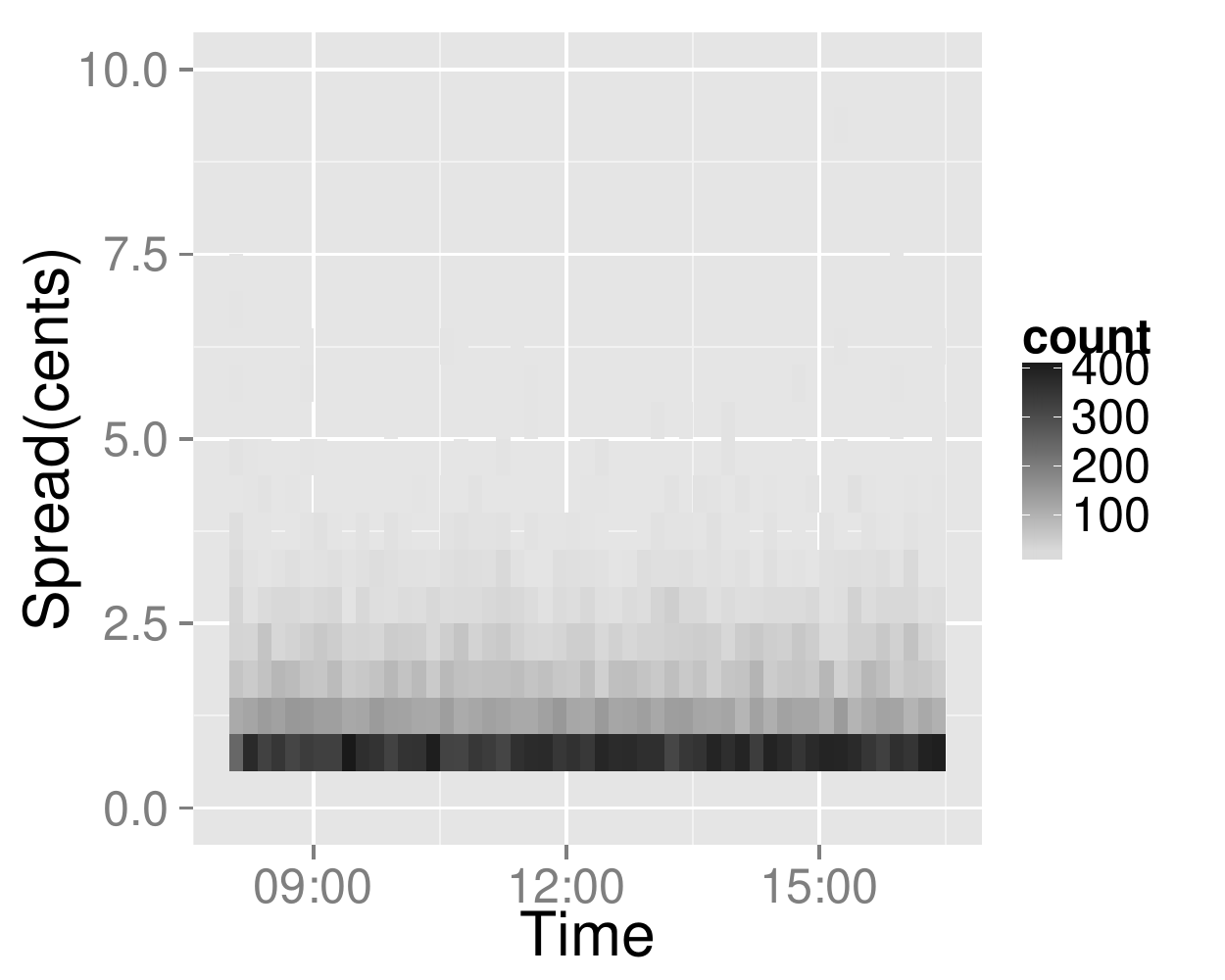}
   \includegraphics[width=0.49\textwidth]{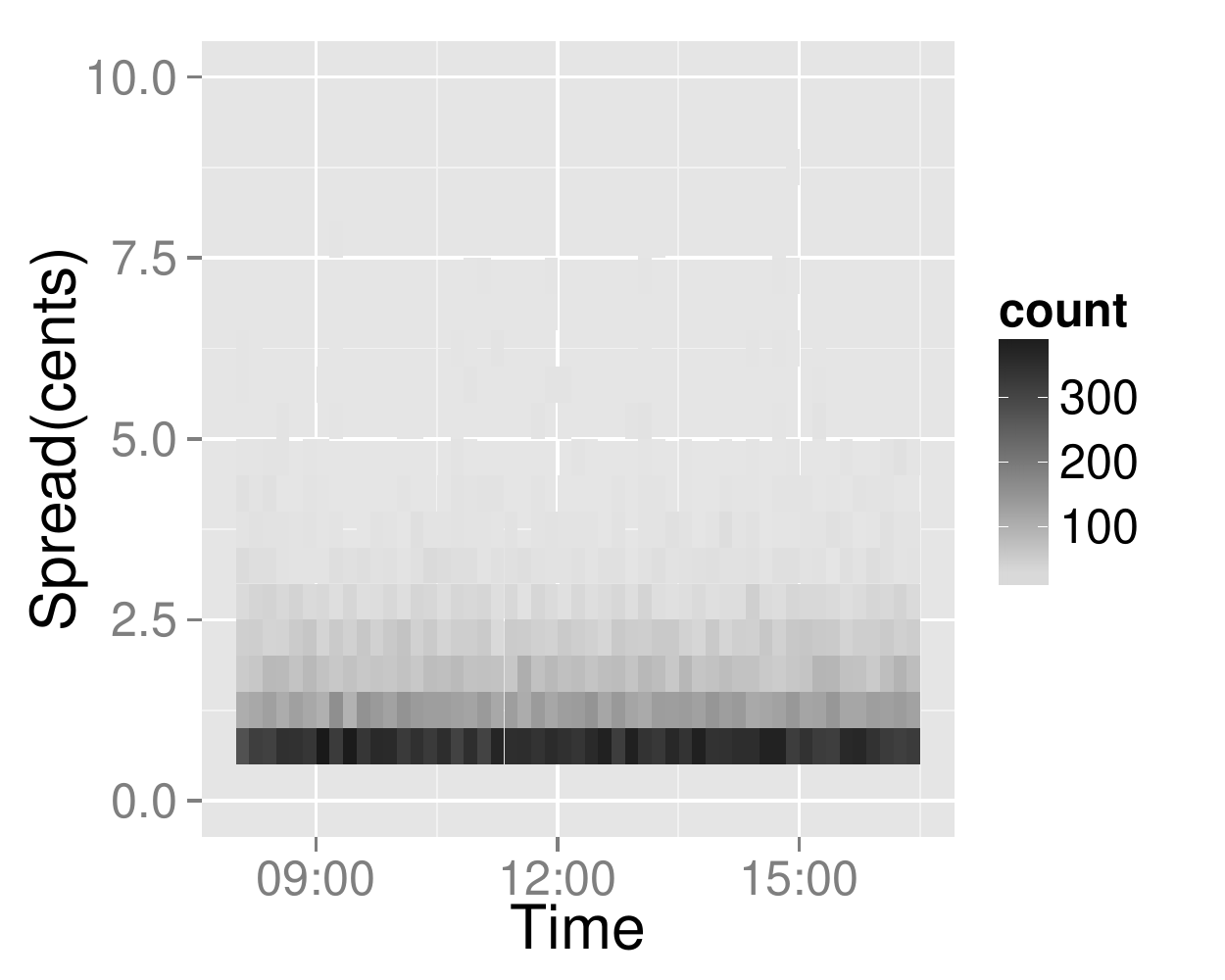}    
 \caption{Heatmaps of the intra-day value of the spread for (Left): The MAP particle from the estimation procedure and (Right): MMSE particle estimates.
 }
 \label{fig:multressimspread}
 \end{center}
 \end{figure}

\begin{figure}[ht]
 \begin{center}
    \includegraphics[width=0.4\textwidth]{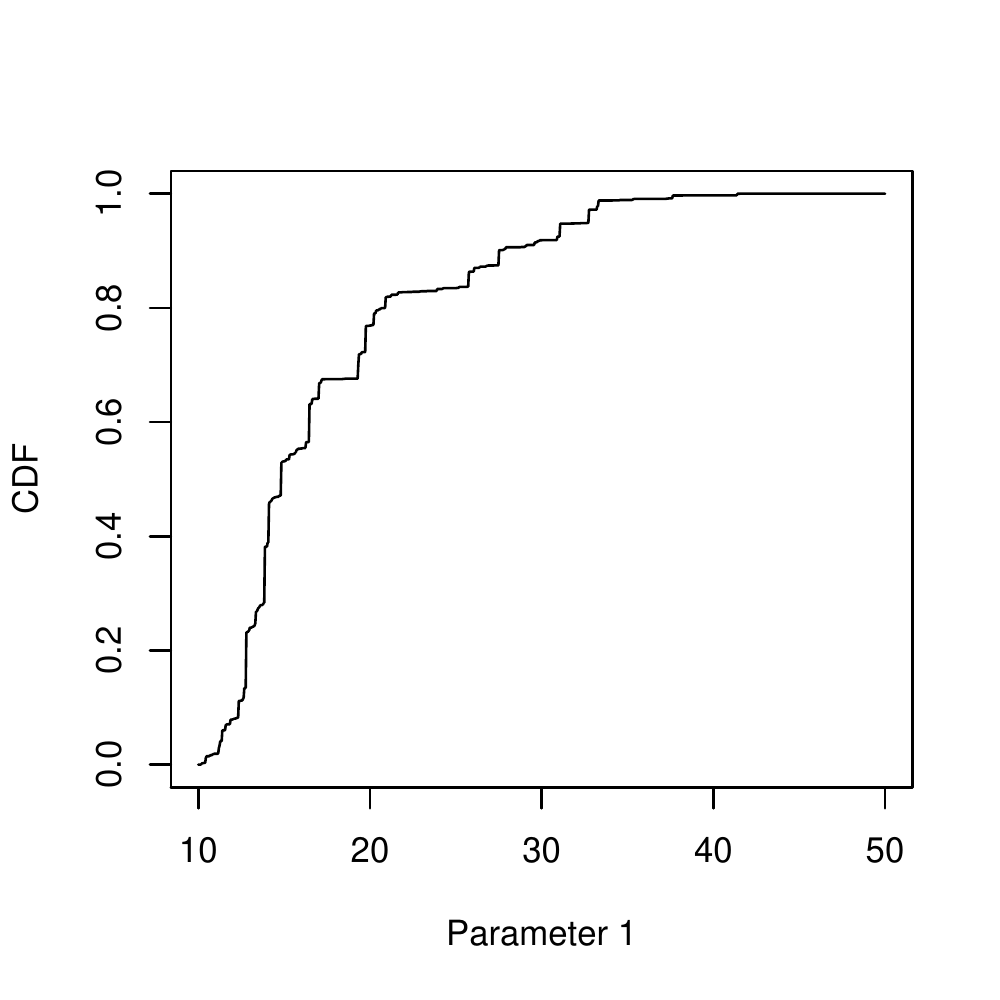}
    \includegraphics[width=0.4\textwidth]{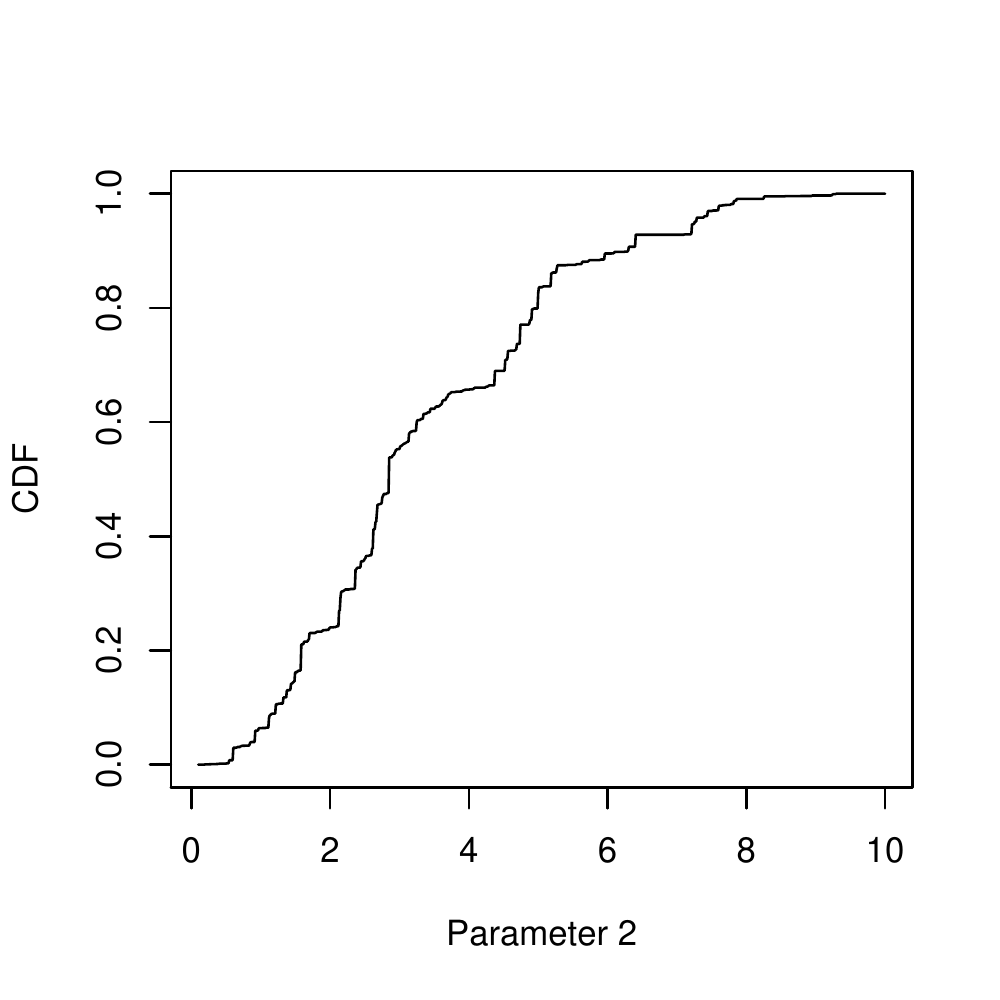}
    \includegraphics[width=0.4\textwidth]{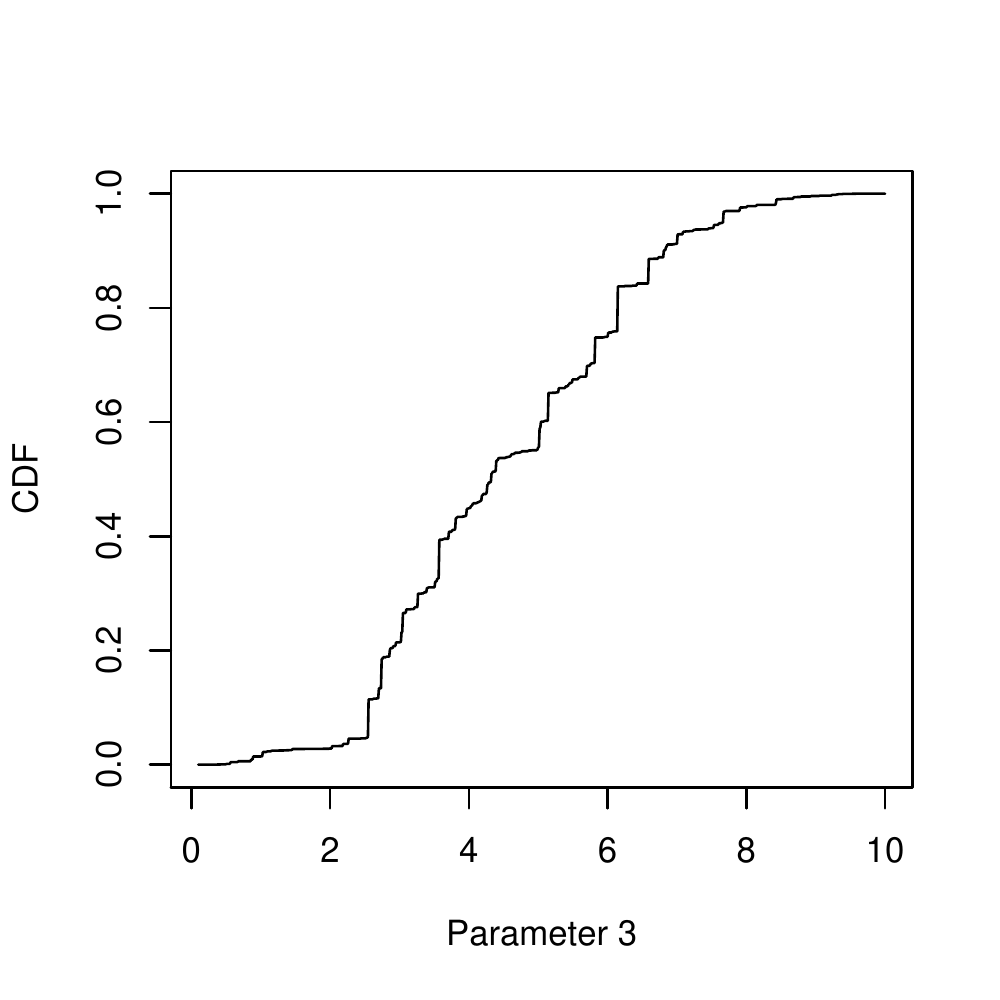}
    \includegraphics[width=0.4\textwidth]{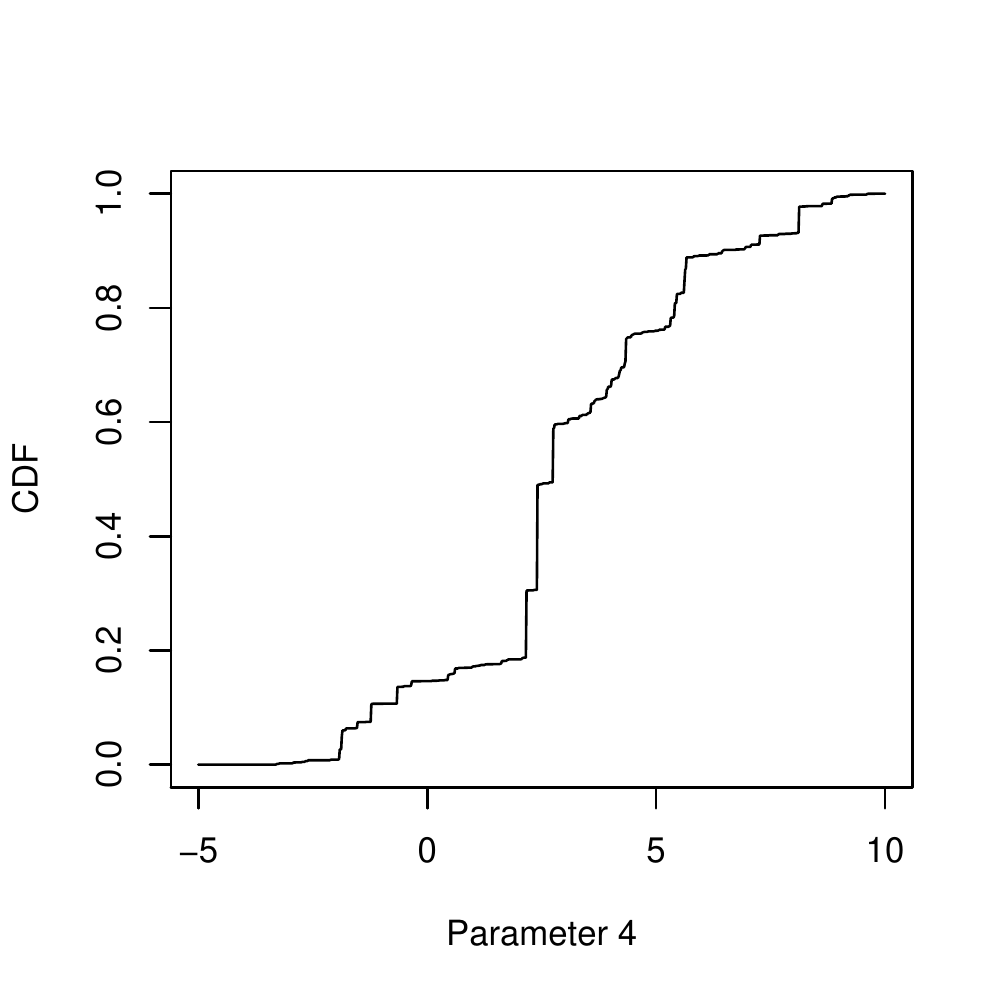}
    \includegraphics[width=0.4\textwidth]{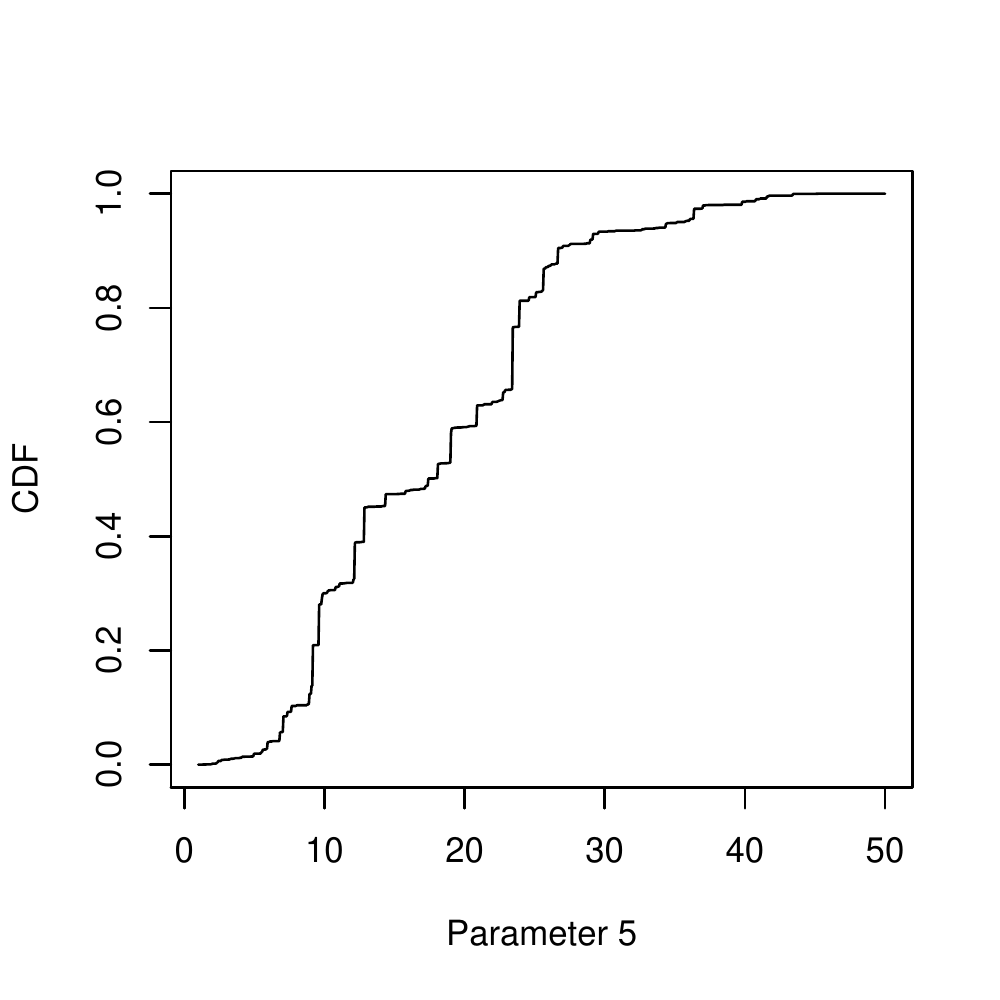}
    \includegraphics[width=0.4\textwidth]{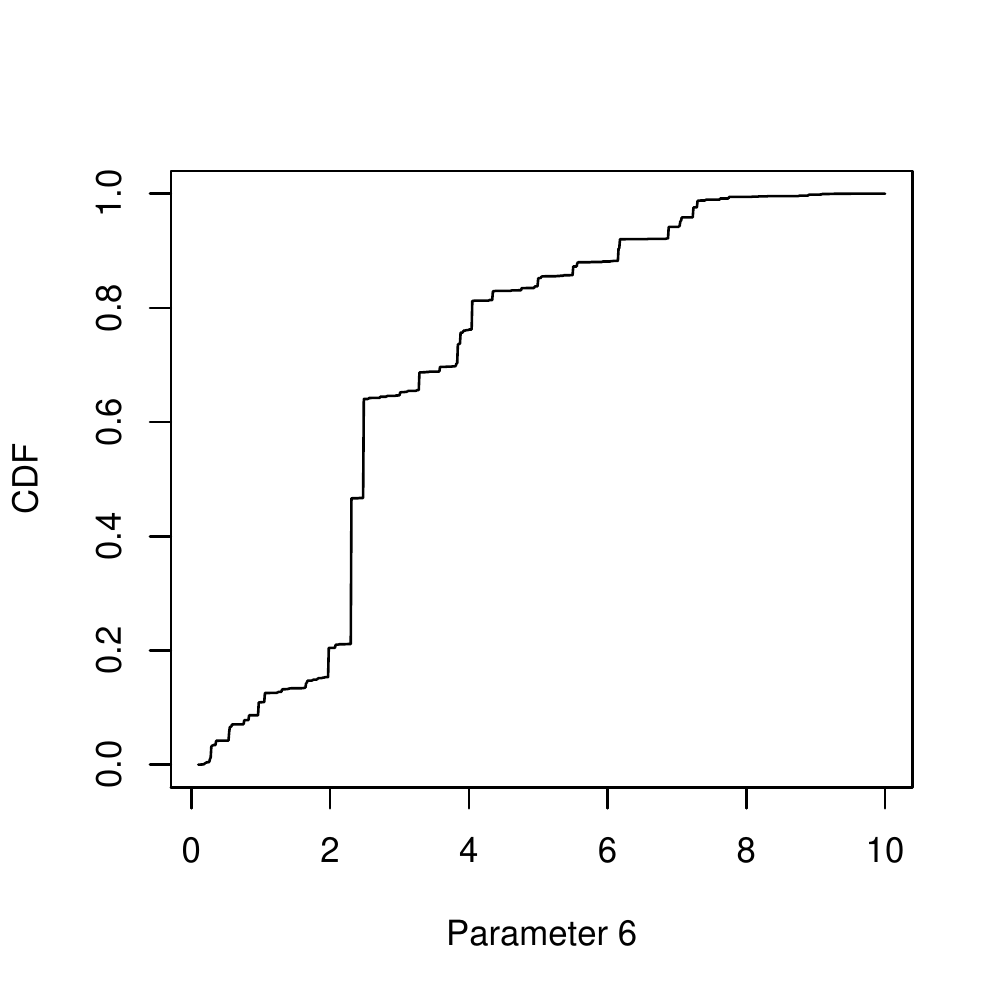}
 \caption{Median of the CDFs for every iteration of the estimation procedure for the parameters of the model. In the figures, parameters 1 to 6 correspond to $\left \{ \mu_0^{LO,p},\mu_0^{LO,d}, \mu_0^{MO},\gamma_0,\nu, \sigma^{MO} \right \}$, respectively.}
 \label{fig:paramdist}
 \end{center}
 \end{figure}

To complete the analysis we also illustrate the median of the resulting marginal posterior distributions for the model parameters obtained from 20 independent runs of the SMC Sampler-ABC algorithm for the BNP Paribas data on 05/03/2012. These results are presented in Figure \ref{fig:paramdist}.

\subsection{Results Comparison to MOEA-II procedure}
   \label{sec:moeaproc}     

The method introduced in \cite{panayi2015stochastic} is a combination of simulation-based indirect inference (II) and multi-objective optimisation, denoted the Multi-objective-II estimation framework. In common with ABC, II is used when one cannot write down the likelihood of the data generating model in closed form, but realisations are easily obtained via simulation given model parameters $\bm{\theta}$. II introduces a new, `auxiliary' model (with parameter vector $\bm{\beta}$), which is fit to a transform of both the real and simulated data ($\bm{y}$ and $\bm{y}^{\ast}(\bm{\theta})$, respectively) and the objective is to find the model parameter vector $\hat{\bm{\theta}}$ which minimises some distance metric $D(\bm{\beta}(\bm{y}),\bm{\beta}(\bm{y}^{\ast}(\bm{\theta}))$.

The multi-objective extension to the standard II procedure pertains to the objective function $D(\bm{\beta}(\bm{y}),\bm{\beta}(\bm{y}^{\ast}(\bm{\theta}))$. Where standard II procedures consider a scalar output of the objective function, the Multi-objective-II method considers a vector-valued output, where each element of the vector pertains to a different feature of the LOB stochastic process. In this framework, the search is then for \textit{non-dominated} parameter vectors, i.e. such that there is no parameter vector in the search space that can unilaterally improve a single criterion (objective function element) without worsening another criterion. The procedure uses the same mutation and crossover kernels outlined in Section \ref{sec:mutker} and outputs a set of \textit{Pareto optimal} solutions, see details in \cite{panayi2015stochastic}.

Where the SMC-ABC method returns a family of particles and associated weights, the MOEA-II procedure returns a family of particles and their non-domination rank. Our comparison is then between the highest weighted particles returned from the former procedure, and the non-dominated particles returned from the latter. The results have been found to be comparable between the two methods, both in terms of achieving similar objective function values and in terms of producing simulations which resemble real financial markets. That is, while not all highest weight/non-dominated particles will give rise to realistic financial market simulations, there is a subset that do.   

While we have tried to provide a fair comparison between the two methods by utilising the same mutation and crossover operators, we should highlight some differences between the MOEA-II procedure and the SMC-ABC procedure presented in this paper. Firstly, the MOEA-II procedure did not suffer from particle degeneracy when using the adaptive mutation kernel for the covariance mutation operator, and thus the operator in Section \ref{sec:mutker} was utilised as described. Secondly, where the probability of crossover between particles in the MOEA-II procedure was set at the default value of 0.7 in every iteration, this was found to cause additional particle degeneracy issues and thus the probability was reduced to 0.05 (with the additional exclusion of crossing with identical particles described at the front of this section).  

\section{Conclusion}

This chapter has proposed a stochastic agent based liquidity supply and demand based simulation based model to characterize the LOB for an asset traded on an electronic exchange. The calibration of this model to real market LOB data has been performed via a posterior inference procedure that adopted an ABC structure due to the complexity of writing down the resulting likelihood for the LOB agent simulation model. The estimation of the posterior distribution was then shown how to be performed via an adaptive SMC Sampler-ABC algorithm. The results were tested on real data and compared to an indirect inference procedure with multi-objective optimization features.

\newpage
\bibliographystyle{plainnat}
\bibliography{ChapterSMCABC.bib}

\end{document}